\documentclass[acmsmall,screen]{acmart}

\AtBeginDocument{%
  }

\setcopyright{acmlicensed}
\acmDOI{XXXXXXX.XXXXXXX}

\acmConference[Conference 2026]{Proceedings of International Conference}{XXXXX}{XXXX, XXXX, XXXXX}

\usepackage{subcaption}
\usepackage{makecell}
\usepackage{enumitem}
\usepackage{multirow}
\usepackage{color}
\usepackage{colortbl}
\usepackage{listings}
\usepackage{xcolor}
\usepackage{wrapfig}

\usepackage{algorithm}
\usepackage{algpseudocode}

\algdef{SE}[VARIABLES]{Variables}{EndVariables}
   {\algorithmicvariables}
   {\algorithmicend\ \algorithmicvariables}
\algnewcommand{\algorithmicvariables}{\textbf{global variables}}
\algnewcommand{\LineComment}[1]{\State \(\triangleright\) \textit{#1}}

\usepackage{tikz}

\newcommand{\mfoe}{MFOE}
\newcommand{\techname}{MetaFOE}
\newcommand{\techphaseA}{MRG} 
\newcommand{\techphaseB}{FOI} 

\newcommand{\ossfuzz}{OSS-Fuzz}
\newcommand{\tocheck}[1]{\textcolor{black}{#1}}

\newcommand{\nprompt}{\tocheck{five}}

\newcommand{\fuzzhours}{\tocheck{three}}

\newcommand{\nllm}{\tocheck{three}}

\newcommand{\nossfuzzprojects}{\tocheck{127}}
\newcommand{\nossfuzzdrivers}{\tocheck{1,068}} 
\newcommand{\nmfoesubjects}{10}
\newcommand{\ngenMRs}{\tocheck{20}}
\newcommand{\ngenMRsEach}{\tocheck{10}}
\newcommand{\nvmd}{\tocheck{6,228}}
\newcommand{\nassertabort}{\tocheck{367}}
\newcommand{\ntokens}{\tocheck{213 M}}

\newcommand{\ncpuhours}{\tocheck{18,684}}
\newcommand{\ncpuyears}{\tocheck{two}} 

\newcommand{\rqtitle}[1]{RQ#1}
\newcommand{\rqline}[2]{\textbf{\rqtitle{#1}:} #2}

\newcommand{\rangeline}{--}
\newcommand{\placeholder}{\textcolor{red}{XXX}}

\newcommand{\lineno}[1]{(#1)}

\newcommand{\tuple}[1]{\langle#1\rangle}

\makeatletter
    
  \renewcommand*\sectionautorefname{\S\@gobble}
  \renewcommand*\subsectionautorefname{\S\@gobble}
  \renewcommand*\subsubsectionautorefname{\S\@gobble}
\makeatother

\usepackage{tcolorbox}
\newtcolorbox{researchquestionbox}[1][]{
  colback=white,
  colframe=black,
  arc=2.5pt,
  boxrule=1.2pt,
  left=2pt,
  right=2pt,
  top=3pt,
  bottom=3pt,
  boxsep=2pt,
  before skip=6pt,
  after skip=6pt,
  #1
}

\newcommand{\rqbox}[2]{
  \begin{researchquestionbox}
    \textbf{RQ#1:} #2
  \end{researchquestionbox}
}

\newtcolorbox{answerbox}[1][]{
  colback=gray!15,
  colframe=black,
  arc=2.5pt,
  boxrule=1.2pt,
  left=2pt,
  right=2pt,
  top=3pt,
  bottom=3pt,
  boxsep=2pt,
  before skip=10pt,
  after skip=10pt,
  #1
}

\newcommand{\findingbox}[2]{
  \begin{answerbox}
    \textbf{Finding #1:} #2
  \end{answerbox}
}

\begin{document}

\title{Investigating Metamorphic Fuzz Oracle Enhancement via Large Language Models}

\author{Ruixiang Qian}
\email{qianrx@smail.nju.edu.cn}
\orcid{0009-0003-5040-3123}
\affiliation{%
  \institution{State Key Laboratory for Novel Software Technology}
  \city{Nanjing University}
  \country{China}
}


\author{Ding Yang}
\email{dingyang@smail.nju.edu.cn}
\orcid{0009-0005-9684-2046}
\affiliation{%
  \institution{State Key Laboratory for Novel Software Technology}
  \city{Nanjing University}
  \country{China}
}

\author{Zengxu Chen}
\email{522025320022@smail.nju.edu.cn}
\orcid{0009-0003-7924-1295}
\affiliation{%
  \institution{State Key Laboratory for Novel Software Technology}
  \city{Nanjing University}
  \country{China}
}

\author{Yuxuan Gao}
\email{522025320035@smail.nju.edu.cn}
\orcid{0009-0006-9718-5460}
\affiliation{%
  \institution{State Key Laboratory for Novel Software Technology}
  \city{Nanjing University}
  \country{China}
}

\author{Chunrong Fang}
\authornote{Chunrong Fang and Zhenyu Chen are the corresponding authors.}
\email{fangchunrong@nju.edu.cn}
\orcid{0000-0002-9930-7111}
\affiliation{%
  \institution{State Key Laboratory for Novel Software Technology}
  \city{Nanjing University}
  \country{China}
}


\author{Chao Zhang}
\email{chaoz@tsinghua.edu.cn}
\orcid{0000-0001-7894-8828}
\affiliation{%
  \institution{Tsinghua University}
  \city{Beijing}
  \country{China}
}

\author{Zhenyu Chen}
\email{zychen@nju.edu.cn}
\authornotemark[1]
\orcid{0000-0002-9592-7022}
\affiliation{%
  \institution{State Key Laboratory for Novel Software Technology}
  \city{Nanjing University}
  \country{China}
}


\begin{abstract}
Fuzz drivers are important components of greybox fuzzing; they encapsulate the target interface of the tested library, define the test space, and largely determine the outcomes of fuzzing.
Fuzz drivers rely on (fuzz) oracles to detect potential bugs.
As a common practice, existing fuzz drivers typically focus on security testing and adopt program crashes as the oracle. 
Although general and practical, this practice overlooks the functionality of the library under test, thereby restricting the bug-mining capability of greybox fuzzing.

In this paper, we present the first study on metamorphic-based fuzz oracle enhancement (\mfoe{}), which aims to improve existing fuzz drivers by incorporating metamorphic-based oracles.
The metamorphic-based oracles are grounded on metamorphic relations (MRs), 
which reflect the expected relationships between multiple inputs and their corresponding outputs.
However, both the construction and integration of metamorphic-based oracles require a deep understanding of the fuzz target, making automatic \mfoe{} particularly challenging.
Modern large language models (LLMs), with their advanced capabilities in code understanding and generation, provide a promising opportunity to overcome this obstacle, inspiring us to develop an LLM-based framework.

We name the LLM-based \mfoe{} framework as \techname{} and extensively evaluate it on fuzz drivers selected from \ossfuzz{}.
To conduct a comprehensive investigation, we adopt \nllm{} modern LLMs and \nprompt{} prompt strategies to configure \techname{}.
In total, we generate \tocheck{3,475} MRs, \tocheck{77.3\%} of which are applicable.
Building on these MRs, we implement \tocheck{12,351} meta drivers (i.e., drivers incorporating metamorphic-based oracles), of which \tocheck{6,228} are valid.
After \fuzzhours{} hours of fuzzing, these valid meta drivers achieve an average \tocheck{18.7\%} improvement in edge coverage and triggered \tocheck{1,528} unique crashes.
Our study highlights the necessity of incorporating fuzz drivers with metamorphic-based oracles and demonstrates the feasibility of an LLM-based automatic \mfoe{}, offering valuable insight to the fuzzing community.  
\end{abstract}




\maketitle

\section{Introduction}
\label{sec:intro}

Greybox fuzzing is a simple yet super effective automatic testing technique \cite{marcel2025software,zhu2022fuzzing}.
With a favorable trade-off between effectiveness and efficiency, greybox fuzzing has become an important testing capability, empowering notable bug-hunting services \cite{serebryany2017oss,syzbot} and leading to the discovery of tens of thousands of software vulnerabilities \cite{osvdev,afltrophy}.
The fuzz driver (or fuzz harness\footnote{The terms ``fuzz driver'' and ``fuzz harness'' are both widely adopted in the fuzzing community and are interchangeable with each other. In this paper, we primarily use the term fuzz driver.}) is a vital component of greybox fuzzing, especially for domains such as library testing \cite{libfuzzertarget,zhang2024effective}. 
Typically, a fuzz driver functions as the entry point for test executions; it not only encapsulates the units to be tested (e.g., library APIs) but also delimits the reachable code regions, which largely determine the results of fuzzing \cite{lyu2024prompt,galland2025invivo}.
Substantial manual efforts have been invested in developing and maintaining fuzz drivers \cite{ossfuzz,gorz2025empirical}, and various automated techniques have been proposed along this line of research \cite{jeong2023utopia,babic2019fudge,chen2023hopper,toffalini2025liberating,liu2025promefuzz}.
All of these advances underscore the importance of crafting high-quality fuzz drivers.

Although numerous works have been devoted to driver generation, they mainly concentrate on the synthesis of test sequences (i.e., ordered and structured API calls \cite{sherman2025no}) while largely overlooking the construction of \textit{fuzz oracles}.
In software testing, an \textit{oracle} is the mechanism used to determine whether an observed program behavior is a bug \cite{barr2014oracle}; and we refer to those employed in fuzzing as fuzz oracles \cite{manes2018fuzzing}.
To understand the current status of the fuzz oracles used in existing drivers, we conduct a systematic inspection (detailed results are in \autoref{subsec:rq-existing-oracles}) of the source code of \nossfuzzdrivers{} drivers from \ossfuzz{} \cite{ossfuzz}.
We find that 63.9\% of the drivers \textit{do not} implement any dedicated bug-detection mechanisms and rely solely on program crashes to indicate bugs.
Among the remaining drivers, 50.1\% only implement ``pseudo oracles'' to check the validity of test inputs and the drivers themselves.
Only 18.9\% of the inspected drivers implement functional oracles that target the libraries under test, highlighting the critical need to improve existing fuzz oracles.

In this paper, we conduct the first study investigating the enhancement of fuzz oracles, in particular those embedded in existing drivers.
In the prior art, researchers have recognized the shortcomings of fuzz oracles and have proposed novel techniques to tackle this problem \cite{zhang2022debloating,serebryany2012addresssanitizer,zhang2021sanrazor,han2020metamorphic}. 
As a powerful technique for alleviating the oracle problem \cite{rigger2020testing,guo2024cootest,mu2025improving,segura2018metamorphic, liu2013effectively}, metamorphic testing has naturally attracted the attention of the fuzzing community and has been adopted to enhance fuzz oracles.
For example, \citet{pan2024edefuzz} summarize the characteristics of excessive data exposures (EDEs) and design novel oracles to fuzz and expose EDEs in real-world web applications.
\citet{kokkonis2025rosa} identify the symptoms of code-level backdoors and construct a metamorphic (test) oracle to assist AFL++ \cite{afl++} in detecting run-time backdoor triggers.
These works (which can be generally termed \textit{metamorphic fuzzing}) have proposed insightful solutions to the fuzz oracle problem, yet only concentrate on constructing oracles to expose specific types of bugs (i.e., EDEs and code-level backdoors) and do not address the lack of well-crafted oracles in existing drivers, which motivates us to explore this issue.

Inspired by the success of metamorphic fuzzing, we propose improving existing fuzz drivers with metamorphic oracles.
We refer to this pipeline as metamorphic-based fuzz oracle enhancement (\mfoe{}) and identify two essential tasks of it: 
\begin{enumerate}[leftmargin=*,topsep=3pt]
    \item \textbf{Metamorphic Relation Generation (\techphaseA{})}. 
    Analyzing fuzzed libraries to extract metamorphic relations (MRs) \cite{chen2018metamorphic}, which are the core of metamorphic testing and the foundation for constructing metamorphic oracles.
    \item \textbf{Fuzz Oracle Implementation (\techphaseB{})}.
    Analyzing given fuzz drivers, matching them with the generated MRs, and implementing appropriate oracles to construct enhanced drivers.
\end{enumerate}
Both of \techphaseA{} and \techphaseB{} require a deep understanding of the input software assets (i.e., fuzzed libraries and fuzz drivers), which poses a tricky obstacle to \mfoe{}.
In recent years, modern large language models (LLMs) have advanced substantially and shown strong capabilities in code generation and understanding \cite{sun2025source,shang2025large,AAAI26-incoherence}, which opens opportunities to mitigate the obstacle to \mfoe{}.
Moreover, pioneer researchers have demonstrated that LLMs can be applied to the generation of fuzz drivers  \cite{zhang2024effective,lyu2024prompt} and MRs \cite{shin2024towards,zhang2025can}, which inspires us to develop an LLM-based workflow for \mfoe{}. 

We name the LLM-based \mfoe{} pipeline as \techname{} and conduct extensive experiments to gain an empirical understanding of it.
Specifically, we systematically evaluate the capabilities of LLMs in \lineno{1} generating valid MRs for given fuzz drivers and target libraries, and \lineno{2} constructing metamorphic oracles based on valid MRs and improving given fuzz drivers;
in this evaluation, \nllm{} representative LLMs are adopted, and \nprompt{} mainstream prompting strategies are investigated.
For each generated MR, we manually validate it following pioneering work \cite{zhang2025can}.
For each enhanced driver candidate produced by an LLM, we verify its validity through a three-phase process with checks at both the compilation and runtime levels.

With \nmfoesubjects{} original fuzz drivers and the corresponding project documents as raw input, \techname{} generates \tocheck{3,475} MRs and \tocheck{12,351} meta drivers in total, of which \tocheck{77.3\%} and \tocheck{50.4\%} are applicable and valid. 
For the \tocheck{6,228} valid meta drivers, we further evaluate their fuzzing capabilities (i.e., code coverage and crash detection) through \fuzzhours{}-hour campaigns. 
The results show that the meta drivers achieve \tocheck{18.7\%} more edge coverage and trigger \tocheck{1,528} unique crashes.
These results reveal that metamorphic oracles can not only improve code coverage but also crash detection.
We also experiment \techname{} using \nllm{} LLMs and \nprompt{} prompt strategies to evaluate its sensitivity to different configurations.
The main contributions of this paper are as follows:
\begin{itemize}[topsep=3pt, leftmargin=*]
    \item \textbf{Novel Pipeline}.
    We propose a novel task named metamorphic-based fuzz oracle enhancement, and develop an LLM-based pipeline \techname{} to realize it.
    To the best of our knowledge, this is the first work to improve fuzz oracles encoded in existing drivers.
    \item \textbf{Systematic Study}.
    We conduct a systematic study of existing fuzz oracles by inspecting \nossfuzzdrivers{} drivers from \ossfuzz{}.
    We also perform extensive experiments to evaluate \lineno{1} \techname{}'s performance in \techphaseA{} and \techphaseB{}, which consume \ntokens{} tokens; and \lineno{2} the fuzzing capability of valid meta drivers, which spends more than \ncpuyears{} CPU years.
    \item \textbf{Enhanced Drivers}.
    Through extensive experiments, \techname{} generates \nvmd{} valid meta drivers, which cover \tocheck{18.7\%} more edges than original drivers on average, and trigger \tocheck{1,528} crashes.
    We release these generated meta drivers in our artifact \cite{techArtifact}.
\end{itemize}

\section{Background}
\label{sec:bk}
This section provides background knowledge on metamorphic-based fuzz oracle enhancement, including greybox fuzzing, metamorphic testing, and the large language model.

\subsection{Greybox Fuzzing}
\label{subsec:gf}
Greybox fuzzing is a representative modern software security technique responsible for numerous bugs. 
Building on the basic fuzzing workflow, greybox fuzzing additionally leverages code coverage as the primary feedback. 
It continuously generates a large number of test inputs to repeatedly exercise fuzz targets (invoked through fuzz drivers), aiming to explore previously unseen code regions or trigger latent bugs via massive executions \cite{funfuzztosem,kokkonis2025rosa}.

\subsubsection*{Fuzz Driver}
\begin{figure*}

    \begin{minipage}[b]{.49\linewidth}
        \centering
        \includegraphics[width=\linewidth]{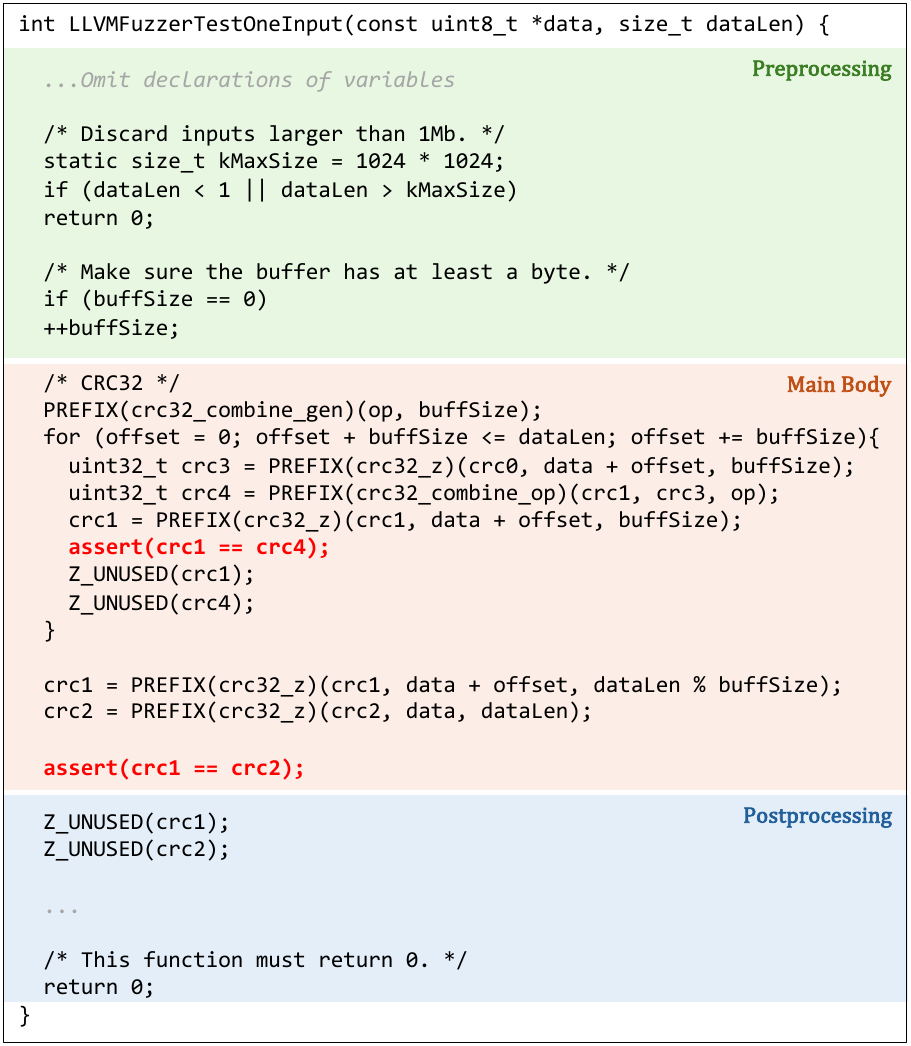}
        \subcaption{
            Driver \texttt{checksum\_fuzzer}  implementing oracles with \texttt{assert} statements \cite{cblosc2checksumfuzzer}. 
        }
        \label{fig:withoracle}
    \end{minipage}
    \begin{minipage}[b]{.49\linewidth}
        \centering
        \includegraphics[width=\linewidth]{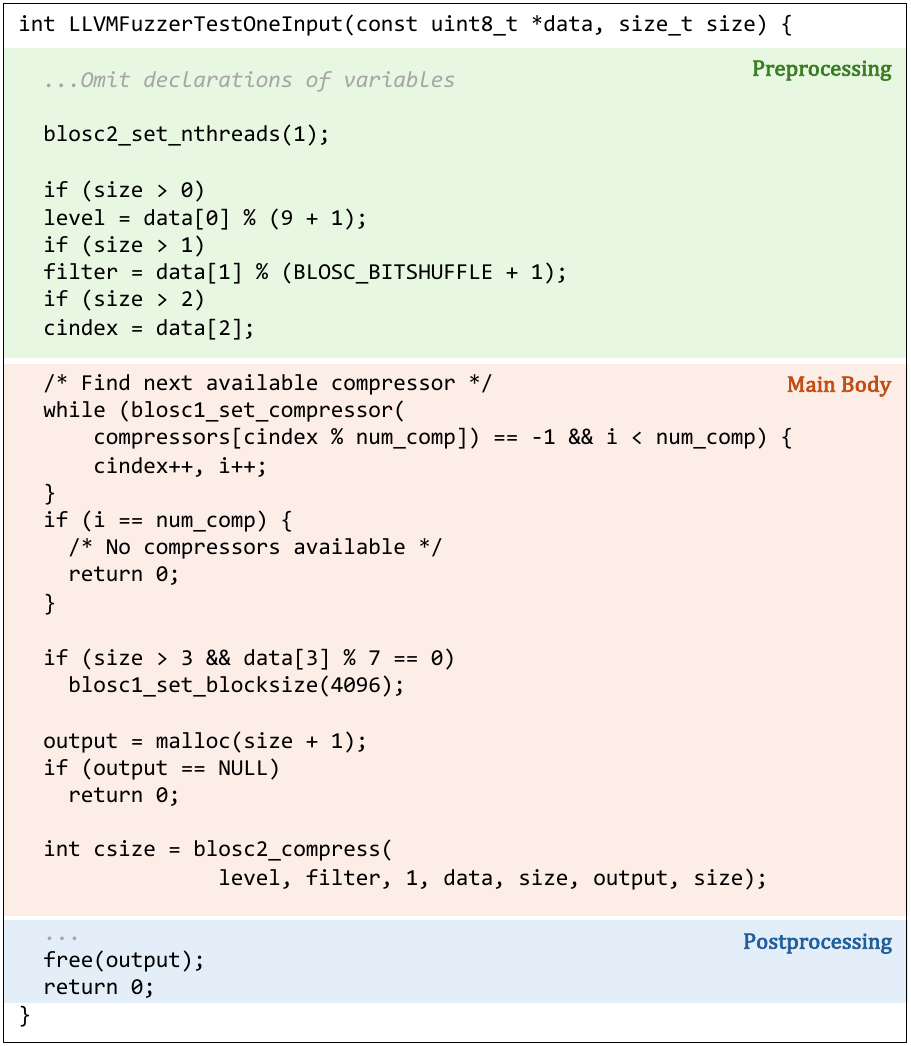}
        \subcaption{
            Driver \texttt{fuzz\_compress\_chunk} implementing no explicit oracles \cite{cblosc2fuzzcompresschunk}.
        }
        \label{fig:withoutoracle}
    \end{minipage}
    \caption{Two LibFuzzer-style fuzz drivers from C-Blosc2, a project integrated in \ossfuzz{}.}
    \label{fig:libfuzzer-target}
    \Description{Two LibFuzzer-style fuzz drivers from C-Blosc2, a project integrated in \ossfuzz{}}
\end{figure*}

The fuzz driver is a critical component of greybox fuzzing that implements the execution entry point of the fuzz target (e.g., the library under test).
In addition to the \textit{main} step that invokes the Abstract Programming Interfaces (APIs) to be fuzzed, a qualified fuzz driver usually includes two other steps:
\lineno{1} a \textit{preprocessing} step to prepare test inputs for the tested interfaces, and \lineno{2} a \textit{ postprocessing} step to release the resources \cite{zhang2024effective, libfuzzertarget}.
\autoref{fig:libfuzzer-target} illustrates two LibFuzzer-style drivers from the C-Blosc2 project; the code blocks that implement the preprocessing, main, and postprocessing steps are colored orange, green, and blue.

In our study, we focus on fuzz oracles and dissect a driver into two parts: the \textit{test sequence} and the  \textit{test oracle}.
Specifically, the test sequence part defines the test logic of the tested APIs (i.e., the invocation order and parameter dependencies), and the test oracle part validates the execution results of the sequence to determine whether a bug is observed.
The driver shown in \autoref{fig:withoracle} encompasses both the test sequence and the test oracle parts, where the explicit oracles\footnote{Following \cite{barr2014oracle}, we distinguish between \textit{explicit oracles}, which explicitly check program properties (e.g., assertions), and \textit{implicit oracles}, which capture properties that should \textit{not} be violated, such as program crashes.} implemented with \texttt{assert} statements are highlighted in red.
C-Blosc2 is a C library for data compression, and the \texttt{assert} oracles embedded in the driver \texttt{checksum\_fuzzer} are designed to verify the equivalence and consistency of CRC32, which are fundamental building blocks in data compression.
With such oracles, a driver can help fuzzing to detect not only plain bugs that lead to program crashes, but also functional bugs that violate the semantic correctness of critical APIs, such as CRC32 implemented in the C-Blosc2 library.

However, unlike the \texttt{assert} oracles illustrated in \autoref{fig:withoracle}, the common practice in fuzzing still largely relies on program crashes (i.e., implicit oracles \cite{barr2014oracle}) as the sole oracle.
As shown in \autoref{fig:withoutoracle}, the driver \texttt{fuzz\_compress\_chunk} does not perform any explicit checks on the execution results; it just constructs inputs, invokes the target APIs (e.g., \texttt{blosc1\_set\_compressor}), and frees the allocated memory, relying on repeated executions of this workflow during fuzzing to expose bugs.
Such implicit oracles are overly simplistic and may miss important bugs, revealing limitations in existing fuzz oracles. 
This observation motivates us to carry out this study.


\subsubsection*{OSS-Fuzz}
\ossfuzz{} is a continuous fuzzing service provided by Google to support open-source projects \cite{ossfuzz,osvdev}.
With incentives in place, an open-source library can choose to participate in \ossfuzz{} for continuous fuzzing.
To enable integration into the \ossfuzz{} infrastructure, library vendors should prepare standard artifacts mandated by \ossfuzz{} (e.g., integration configurations and fuzz drivers) before officially requesting inclusion \cite{ossfuzzbuild}.
To date, \ossfuzz{} has successfully integrated more than 1,000 projects, supporting the fuzzing of libraries written in mainstream languages such as C/C++, Java, and Python \cite{ossfuzzintrospector}.
By aggregating abundant fuzzing assets (e.g., initial seed corpora, fuzz-exposed bugs, and fuzz drivers) from actively maintained open-source projects, \ossfuzz{} has become a valuable research testbed, facilitating a number of studies in areas such as regression fuzzing \cite{zhu2021regression}, CI/CD fuzzing \cite{huang2026directed, sharma2024effective}, and empirical fuzzing \cite{zhang2024effective,ding2021empirical}.
Following this line of research, we adopt \ossfuzz{} as our primary data source and conduct this empirical study with the fuzz oracles integrated in \ossfuzz{}.

\subsection{Metamorphic Testing}
\label{sec:bk-mt}
Metamorphic testing was first proposed by \citet{chen1998metamorphic} to address the test generation problem.
During the past few decades, it has evolved into a prominent remediation of the test oracle problem \cite{chen2018metamorphic}.
Given an algorithm $f$ (i.e., the design specification) and a program $p$ that implements this algorithm, metamorphic testing aims to extract a set of metamorphic relations (MRs) $R=\{\mathcal{R}_1, \mathcal{R}_2,...,\mathcal{R}_m\}$ from $f$ and leverage them to validate whether $p$ is implemented correctly.
Essentially, an MR $\mathcal{R}$ is a necessary property that the algorithm $f$ (and thus its implementation $p$) should satisfy. 
If an input–output pair violates $\mathcal{R}$ on $p$, then the output is deemed erroneous, indicating the presence of bugs in $p$.
Let $x$ denote a concrete input and $f(x)$ denote the corresponding output produced by $f$.
Once an MR $\mathcal{R}=\tuple{x_1,x_2,\ldots,x_n,f(x_1),f(x_2),\ldots,f(x_n)}$ is identified and established, metamorphic testing typically proceeds in four steps to generate test cases and validate the correctness of $p$ \cite{chen2018metamorphic}, as follows:
\begin{itemize}[leftmargin=*, topsep=3pt]
    \item Construct $\mathcal{R}'$ by replacing $f$ in $\mathcal{R}$ with $p$, where $\mathcal{R}'$ is the MR used to validate $p$.
    \item Given a sequence of source test cases $\tuple{x_1, x_2, \ldots, x_k}$, execute the program $p$ on each test case to obtain the corresponding outputs $\tuple{p(x_1), p(x_2), \ldots, p(x_k)}$.
    \item Construct the follow-up test cases $\tuple{x_{k+1}, x_{k+2}, \ldots, x_n}$ according to $\mathcal{R}'$, execute them to obtain the corresponding outputs $\tuple{p(x_{k+1}), p(x_{k+2}), \ldots, p(x_n)}$.
    \item Validate $p$ by checking whether the execution results, i.e., $\tuple{p(x_{k+1}),  p(x_{k+2}), \ldots, p(x_n)}$, satisfy $\mathcal{R}'$; if $\mathcal{R}'$ is violated, then a bug in $p$ is revealed.
\end{itemize}

In this study, we propose enhancing existing fuzz drivers by integrating metamorphic oracles, which are derived and implemented based on MRs.
The programs under test are the libraries invoked by the fuzz drivers, with test generation handled by the fuzzer. 
The MRs should therefore be extracted from both the libraries under test and the associated drivers, and we address this task using large language models (LLMs).
Next, we further dissect the tasks involved in fuzz oracle enhancement in \autoref{sec:task-def}, and elaborate on the proposed LLM-based workflow in \autoref{sec:study-framework}.


\section{Metamorphic-based Fuzz Oracle Enhancement}
\label{sec:task-def}
\newcommand{\fuzzDriver}{\mathcal{D}}
\newcommand{\testSeq}{\mathcal{T}}
\newcommand{\libTest}{\mathcal{L}}
\newcommand{\oracle}{\mathcal{O}}
\newcommand{\oracleImp}{\mathcal{O}_i}
\newcommand{\oracleExp}{\mathcal{O}_e}
\newcommand{\foeTask}{\tau}
\newcommand{\stepMapA}{\rho}
\newcommand{\stepMapB}{\phi}
\newcommand{\stepMapC}{\psi}
\newcommand{\propertySet}{\mathbb{P}}
\newcommand{\property}{\mathcal{P}}
\newcommand{\metaR}{\mathcal{R}}




Modern greybox fuzzing typically relies on the program crash---a type of implicit oracle \cite{barr2014oracle}---to identify bugs.
This practice restricts the range of bugs that can potentially be detected, as crash-only oracles inherently overlook the functional semantics of the fuzz targets, i.e., the libraries bridged by the drivers.
To address this limitation, we propose \textit{metamorphic-based fuzz oracle enhancement} (\mfoe{}), a novel pipeline that augments existing fuzz drivers with metamorphic oracles extracted from the libraries under test.

Let $\fuzzDriver = \tuple{\testSeq, \oracleImp}$ be a fuzz driver constructed according to typical fuzzing practice, where $\oracleImp$ represents an implicit oracle that monitors unintended program crashes.
The goal of \mfoe{} is to modify $\fuzzDriver$ into $\fuzzDriver'=\tuple{\testSeq, \oracleExp}$ by replacing the original $\oracleImp$ with an explicit oracle $\oracleExp$, which is constructed based on a metamorphic relation (MR).
Drawing on the metamorphic testing workflow described in \autoref{sec:bk-mt}, we identify three essential tasks in \mfoe{}, as follows:
\begin{enumerate}[leftmargin=*, topsep=3pt]
    \item \textbf{MR Extraction}. 
    Let $\libTest$ be the library behind the fuzzer driver $\fuzzDriver$.
    The MR extraction task is defined as a mapping $\stepMapA(\libTest) = R$, where $R=\{\metaR_1, \metaR_2, ..., \metaR_n\}$ is the set of extracted MRs. 
    
    \item \textbf{Oracle Instantiation}.
    To function as a proper oracle (e.g., an assertion), an MR needs to be instantiated in the context of the fuzzer driver.
    Let $\fuzzDriver$ be a driver and $\metaR$ be an MR extracted from the library exercised by $\fuzzDriver$. 
    The oracle instantiation task is defined as a mapping $\stepMapB(\fuzzDriver, \metaR) = \oracleExp$ that crafts and outputs a metamorphic-based explicit oracle $\oracleExp$. 

    \item \textbf{Oracle Integration}.
    The last step of \mfoe{} is to integrate the constructed oracle $\oracleExp$ into the original driver $\fuzzDriver$.
    As such, the oracle integration task is defined as a mapping $\stepMapC(\fuzzDriver, \oracleExp) = \fuzzDriver'$, where $\fuzzDriver' = \tuple{\testSeq, \oracleExp}$ is the modified driver that incorporates a metamorphic-based oracle $\oracleExp$.
    In the rest of the paper, we refer to $\fuzzDriver'$ as the \textit{meta driver} since it is modified based on MRs.    
    
\end{enumerate}

\subsubsection*{An LLM-based Implementation of \mfoe{}}
All three tasks outlined above require a deep understanding of the fuzzed libraries and the original drivers, posing significant challenges to implementing \mfoe{}. 
As LLMs have demonstrated extraordinary capabilities in code-relevant tasks \cite{sun2025source,chen2025security,liu2025promefuzz,shin2024towards,zhang2024autocoderover,ni2025legofuzz}, we are motivated to leverage LLMs to address challenges in \mfoe{}.  
Specifically, we develop \techname{}, an LLM-based framework that implements FOE in two phases.
The first phase is \textit{metamorphic relation generation} (\techphaseA{}), which implements the mapping $\stepMapA$ defined in the MR extraction task.
The second phase is \textit{fuzz oracle implementation} (\techphaseB{}), which implements both the $\stepMapB$ and $\stepMapC$ mappings defined in the oracle instantiation and integration tasks.
The reason for handling $\stepMapB$ and $\stepMapC$ together is that they share $\fuzzDriver$ as input and rely on similar contextual information, such as invoked APIs, relevant variables, and verifiable results.
We further elaborate on \techname{} in \autoref{sec:study-framework}.

\begin{figure*}
    \centering
    \includegraphics[width=\linewidth]{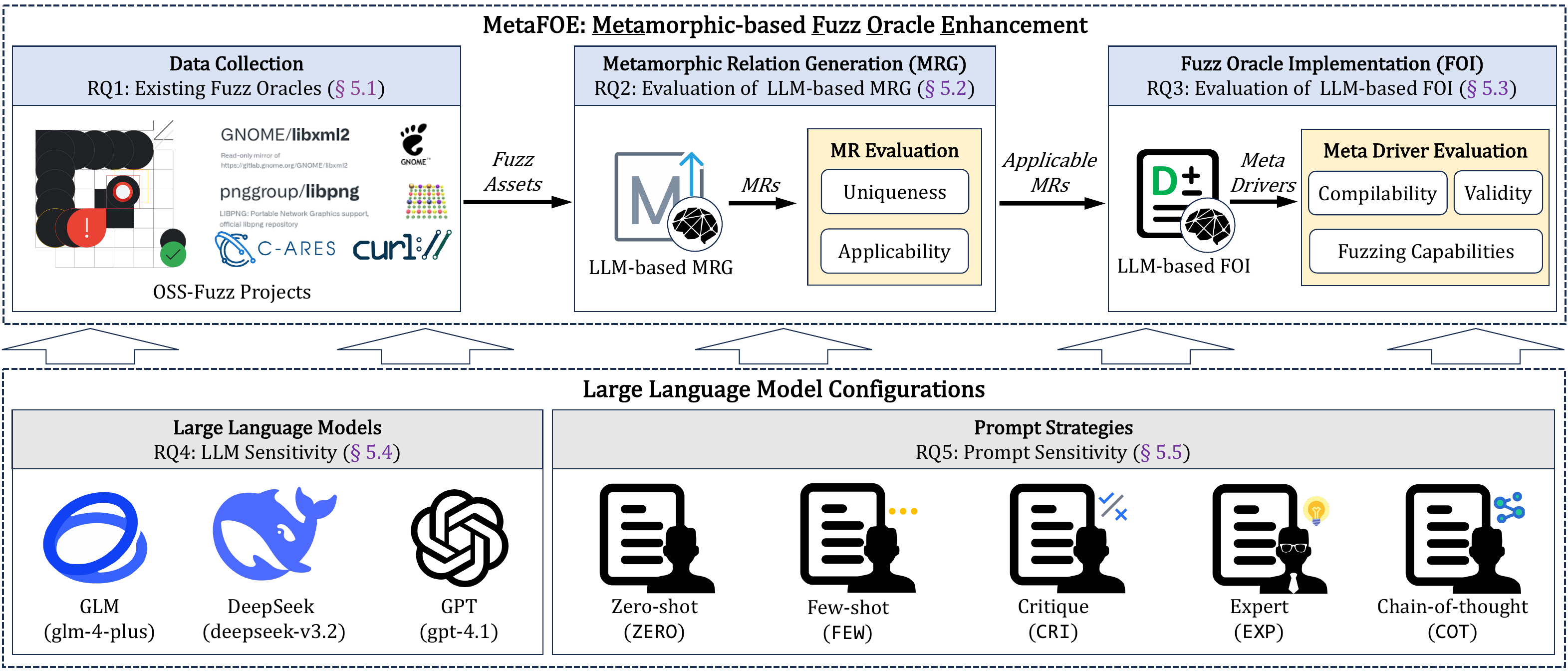}
    \caption{Overview of the study.}
    \label{fig:study-framework}
    \Description{Overview of the study.}
\end{figure*}

\section{Study Design}
\label{sec:study-framework}
\autoref{fig:study-framework} provides an overview of the research questions (RQs), workflows, and subjects (i.e., fuzz drivers, LLMs, and prompt strategies) involved in our study, along with their interrelationships.
Specifically, the upper half of \autoref{fig:study-framework} illustrates the workflow of \techname{}, while the lower half exhibits the LLMs and the prompt strategies included in our study.
Each rectangular frame indicates a working stage (in blue) or a group of study subjects (in gray), where the investigated RQs are presented in the heads of the frames.
Next, we delve into the details of each component.

\subsection{\techname{}}
Taking fuzz assets (e.g., fuzz drivers, seeds, and the source code and documentation of the libraries under test) as input, \techname{} performs two working phases, namely \techphaseA{} and \techphaseB{} (described in \autoref{sec:task-def}), in sequence to modify given fuzz drivers.
During this procedure, a semi-automatic evaluation is conducted to assess the generated MRs in terms of \lineno{1} \textit{uniqueness} and \lineno{2} \textit{applicability}, and an in-process evaluation is conducted to assess the meta drivers along three dimensions: \lineno{1} \textit{compilability}, \lineno{2} \textit{validity}, and \lineno{3} \textit{fuzzing capabilities} (i.e., code coverage and bug discovery). 

\subsubsection{Metamorphic Relation Generation (\techphaseA{})}
In the \techphaseA{} phase, we first prompt an LLM to synthesize MRs, typically producing multiple candidates.
The detailed prompting process for \techphaseA{} is described in \autoref{sec:promptA}.
Since some of the generated MRs are redundant or not suitable for oracle construction \cite{luu2023can,zhang2025can}, we take two steps to evaluate them and discard those that are inapplicable.
To this end, we first deduplicate the MRs using a similarity-based method powered by a small language model\footnote{Similarities are computed using embeddings generated by: \url{https://huggingface.co/sentence-transformers/all-MiniLM-L6-v2}}.
After that, we manually validate the applicability of the MRs following the methodology of pioneering work \cite{luu2023can,zhang2025can}. 
To ensure reliability, the manual validation is carried out by two authors who are experienced in metamorphic testing.
We carefully analyze the features of the libraries under test and validate the corresponding MRs accordingly.
An MR is kept only if both authors agree on its applicability.

\newcommand{\crashratio}{\gamma}
\newcommand{\fpthreshold}{0.01\%}
\newcommand{\compileTrial}{\tocheck{five}}
\newcommand{\shortFuzzMinute}{\tocheck{five}}
\newcommand{\covCamp}{\placeholder{}}
\newcommand{\bugCamp}{\placeholder{}}
\subsubsection{Fuzz Oracle Implementation (\techphaseB{})}
\label{subsubsec:phaseB}

\begin{figure}
    \centering
    \includegraphics[width=\linewidth]{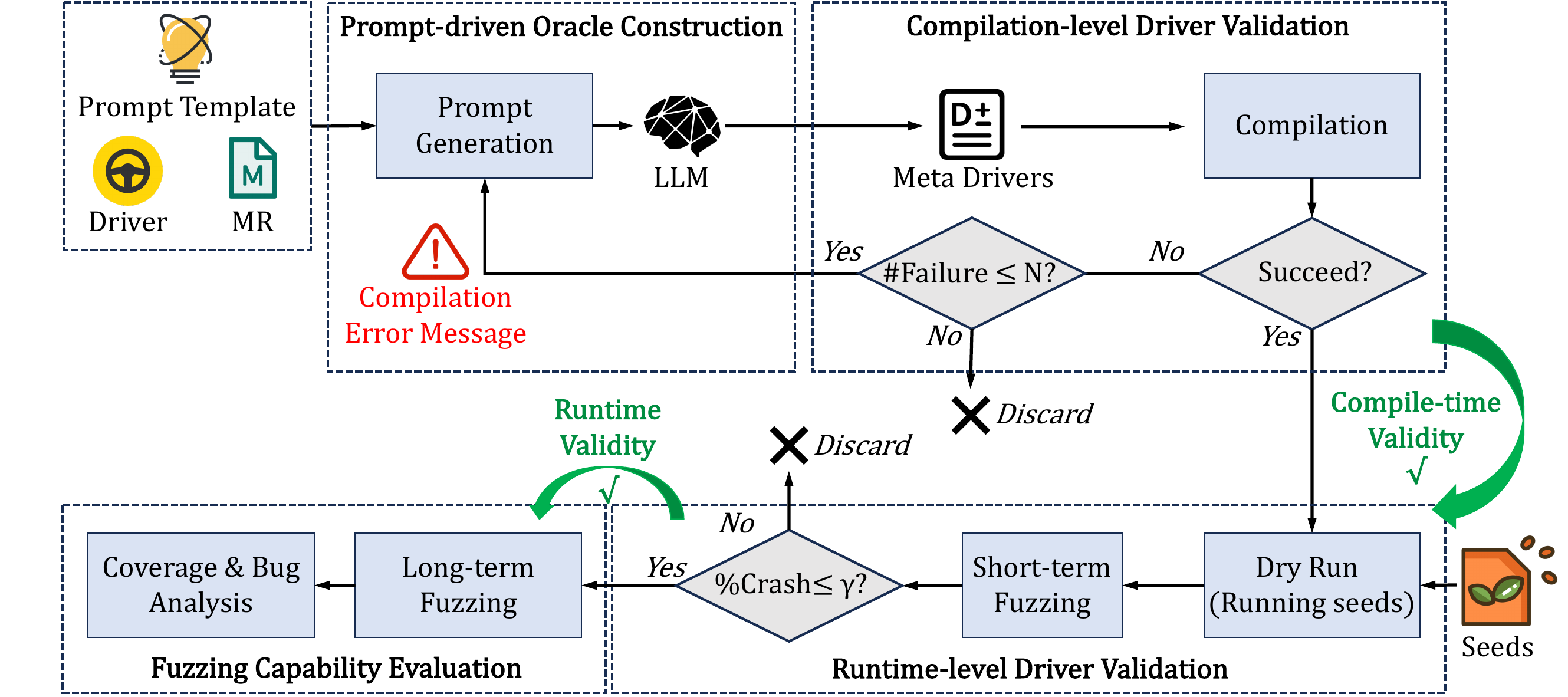}
    \caption{Workflow of the LLM-based \techphaseB{}.}
    \label{fig:phaseB-workflow}
    \Description{Workflow of the LLM-based \techphaseB{}.}
\end{figure}

Based on the MRs retained from the \techphaseA{} phase, the \techphaseB{} phase first modifies the original drivers by implementing and injecting new oracles, and then checks the validity of the modified drivers.
\autoref{fig:phaseB-workflow} provides a closer view of the workflow in \techphaseB{} phase, which comprises four stages:
\begin{enumerate}[leftmargin=*, topsep=3pt]
    \item \textbf{Prompt-driven Oracle Constrution}.
    Given a prompt template, a fuzz driver, and an MR as input, we first craft concrete prompting instructions to elicit the LLM to generate candidate meta drivers. 
    In addition to the input described above, the prompting can also take compilation error messages into account if any candidate meta driver has failed in a previous compilation.
    We describe the prompting process of \techphaseB{} in more detail in \autoref{sec:promptB}.

    \item \textbf{Compilation-level Driver Validation}.
    This stage aims to filter out meta drivers that cannot be compiled.
    As shown in previous studies \cite{gu2024testart, schafer2023empirical, zhang2024effective}, LLMs often struggle to generate compilable code.
    To alleviate this limitation, a widely adopted approach is to iteratively reprompt LLMs to refine the outputs, where compilation error messages are typically used to guide the reprompting.
    Following this approach, we develop a ``reprompt-and-recompile'' iteration to produce as many compilable drivers as possible. 
    We set the number of compilations to at most \compileTrial{} to balance effectiveness and cost.

    \item \textbf{Runtime-level Driver Validation}.
    This stage aims to further filter out meta drivers that are compilable yet invalid at runtime, since a valid meta driver is expected to be not only runnable but also stable.
    After compiling a meta driver, we first run it against seed inputs to ensure that it is runnable.
    A short-term fuzzing (set at \shortFuzzMinute{} minutes) is then conducted to eliminate candidate drivers that \lineno{1} achieve too little coverage, as it often implies that a meta driver implements erroneous behaviors; or \lineno{2} trigger excessive crashes, as it often indicates false positives \cite{zhang2024effective}.
    Specifically, we discard the drivers that \lineno{1} crash in more than \fpthreshold{} of the executions or \lineno{2} achieve less than 10\% of the coverage of the original driver during short-term fuzzing. 
    The coverage of the original driver is the average of five fuzz campaigns.

    \item \textbf{Fuzzing Capability Evaluation}:
    We finally conduct a long-term fuzzing to evaluate how the meta drivers are capable in code coverage and bug detection. 
    For every retained meta driver, we first perform a \fuzzhours{}-hour campaign to obtain raw fuzzing results.
    Based on the results, we then count the coverage achieved by the meta drivers and analyze the triggered crashes.
    
\end{enumerate}

\subsection{Research Questions}
\label{subsec:rqs}









Guided by the roadmap shown in \autoref{fig:study-framework} and centered on the \techname{} framework, we investigate five research questions (RQs), as detailed below.

\rqbox{1 (Existing Fuzz Oracles \autoref{subsec:rq-existing-oracles})}{Do existing fuzz drivers contain oracles? What types of oracles are typically implemented? What are the purposes of the existing oracles?}

Fuzz oracles are critical to the effectiveness of fuzzing; nevertheless, their significance is often overlooked, and their role has not been systematically studied. 
\rqtitle{1} aims to understand the current landscape of fuzzing oracles.   
To this end, we identify the types of existing fuzz oracles and examine them to understand the intended purposes behind their implementations.


\rqbox{2 (Evaluation of LLM-based \techphaseA{} \autoref{subsec:rq-eval-mrg})}{Is LLM-based \techphaseA{} feasible? How well can LLMs construct applicable MRs? To what extent can these MRs be used for oracle construction?}

MRs are the fundamental elements of conducting metamorphic-based fuzz oracle enhancement (\mfoe{}); their quality is crucial to the performance of the produced meta drivers. 
Previous work has explored applying LLMs to synthesize MRs \cite{zhang2025can, luu2023can}, but their use in \mfoe{} remains unexplored.
To bridge this gap, we design RQ2 to evaluate how well LLMs can synthesize applicable MRs, with a particular focus on their suitability for oracle construction.



\rqbox{3 (Evaluation of LLM-based \techphaseB{} \autoref{subsec:rq-driver-enhance})}{Is LLM-based \techphaseB{} feasible? How well can LLMs craft valid meta drivers? Can these meta drivers uncover fresh coverage or detect new bugs?}


The ultimate goal of \mfoe{} is to enhance existing fuzz drivers by leveraging oracles constructed with MRs.
The resulting meta drivers are expected to be stronger than their original version, uncovering additional coverage and exposing potential bugs.
\rqtitle{3} investigates this problem.
We first report statistics on the proportions of meta drivers that achieve the compilation-time and runtime validity, and then evaluate the fuzzing capabilities of the valid meta drivers.

\rqbox{4 (LLM Sensitivity \autoref{subsec:rq-llms})}{Can \techname{} work with different LLMs? How do different LLMs perform on the two tasks, i.e., \techphaseA{} and \techphaseB{}, of \techname{}?}

The LLM is a core building block of \techname{} and can influence the overall performance of \techname{}.
\rqtitle{4} is designed to evaluate the sensitivity of \techname{} to the concrete choice of LLM. 
We aim to understand how different LLMs affect the feasibility and quality of the generated MRs and meta drivers, providing insights into the robustness and generality of the framework.

\rqbox{5 (Prompt Sensitivity \autoref{subsec:rq-prompts})}{Can \techname{} work with different prompt strategies? Will the choice of prompt strategy impacts the results of \techphaseA{} and \techphaseB{}?}


The widespread adoption of LLMs is attributed not only to their strong language processing capabilities, but also to the flexibility of prompting, which requires no extensive data collection or additional training \cite{nong2025appatch}.
Previous studies \cite{sun2025source, zhang2024effective} show that the choice of prompt strategy can significantly affect LLM performance in specific tasks, underscoring the need for a careful prompt design.
Since \techname{} also prompts to tame LLMs, we design \rqtitle{5} to systematically investigate how \techname{} performs with different prompt strategies.

\subsection{Fuzz Drivers}
\label{subsec:data}

All fuzz drivers used in our experiments are collected from \ossfuzz{}.
For \rqtitle{1}, we consider all C/C++ projects and analyze the drivers from those that can be successfully compiled on our local experimental machine.
This process yields \nossfuzzdrivers{} drivers from \nossfuzzprojects{} projects in total; details are described in \autoref{subsec:rq-existing-oracles}.
For \rqtitle{2}\rangeline{}\rqtitle{5}, which systematically evaluate \techname{}, we construct a curated subset by selecting \nmfoesubjects{} projects and choosing one representative driver for each project.
We then run \techname{} with different LLMs and prompt strategies to generate MRs for the selected projects and implement \nvmd{} valid meta drivers.
Detailed statistics are presented in \autoref{subsec:rq-eval-mrg} and \autoref{subsec:rq-driver-enhance}.

\subsection{Large Language Models}
\label{sec:llm}

In this study, we select \nllm{} representative LLMs; their details are as follows.

\textbf{GLM-4-Plus} \cite{glm4repo}.
GLM-4-Plus is a large language model developed by Zhipu AI \cite{du2022glm}. 
It is designed for general-purpose language understanding and generation tasks, including code generation, reasoning, and conversational assistance \cite{glm2024chatglm}. 
Owing to its strong performance on code-related tasks, the GLM series has been widely adopted in both academic research and industrial applications, serving as a core model for software engineering tasks such as fuzzing \cite{fu2024sedar,wang24issre} and software development \cite{du2024evaluating}.

\textbf{DeepSeek-V3.2} \cite{deepseekv32huggingface}.
DeepSeek-V3.2 is a reasoning-oriented LLM developed by DeepSeek AI \cite{liu2025deepseek}; it has exhibited strong multilingual understanding and competitive performance in code generation and completion in multiple programming languages. 
Given its effectiveness in program synthesis, bug detection, refactoring, and documentation generation, the DeepSeek-V3 series has drawn increasing attention in academia, advancing research in intelligent software engineering \cite{zengtypenfuzz, shenICSE26optimization, xia2025demystifying}.



\textbf{GPT-4.1} \cite{gpt41chataccess}.
GPT-4.1 is a recent model in the GPT-4 family developed by OpenAI \cite{gpt41API}. Recognized as one of the most influential LLM families, the GPT series has been widely adopted in both academia and industry, supporting a broad range of applications in software engineering and security \cite{xia2025demystifying,ICSE26-reachabilityGap,zhang2025unlocking,yang2025kernelgpt,yang2024whitefox,meng2024large}. 
As the successor to GPT-4o, GPT-4.1 is more powerful and offers improved coding performance and long-context capabilities; we thus include it in our study.





\subsubsection*{LLM Settings}
In addition to the model and the prompt strategy, temperature is another key factor that can significantly affect LLM outputs \cite{zhang2024effective, xia2025demystifying}. 
To ensure reproducible results while maintaining moderate output variability, we set the temperature to 0.2 for all studied LLMs in both \techphaseA{} and \techphaseB{}.
All other generation parameters are kept at their default values to avoid introducing additional tuning bias.


\subsection{Prompting Method}
\label{subsec:prompt}

In this section, we first describe the prompt strategies adopted in this study and then present the specific prompting process designed for \techphaseA{} and \techphaseB{}.

\subsubsection{Prompt Strategies}
In this paper, we study \nprompt{} prompt strategies, namely zero-shot \cite{kojima2022large}, few-shot \cite{brown2020language}, crit ique \cite{madaan2023self}, expert \cite{ouyang2022training}, and chain-of-thought \cite{wei2022chain}. 
We choose these prompt strategies because they are representative and widely adopted in fuzzing for driver generation \cite{zhang2024effective}, mutator synthesis \cite{ou2024mutators}, and input generation \cite{meng2024large}.
These strategies are also applied to other software engineering tasks, such as metamorphic testing \cite{shin2024towards, zhang2025can}, unit test generation \cite{gu2024testart, cheng2025rug}, and automatic program repair \cite{10.1145/3650212.3680359, zhang2023critical}. 
The details of the studied prompt strategies are as follows.

\newcommand{\zero}{\texttt{ZERO}}
\newcommand{\few}{\texttt{FEW}}
\newcommand{\critique}{\texttt{CRIT}}
\newcommand{\expert}{\texttt{EXP}}
\newcommand{\chain}{\texttt{COT}}

\textbf{Zero-shot (\zero{}).} 
The zero-shot prompt directly instructs an LLM to perform a task without any task-specific examples, relying solely on its pre-trained knowledge.
In this study, we design \zero{} as simple instructions to serve as a basic baseline, following the setup of prior studies that investigate LLMs \cite{sun2025source}.

\textbf{Few-shot (\few{}).} 
In contrast to \zero{}, which provides no task demonstrations in the instructions, the few-shot prompt provides a handful of examples to help the model better understand the task context \cite{geng2024large}.
The examples profile the expected input-output pairs, allowing LLMs to capture task intuitions more accurately.
We include \few{} to investigate whether LLMs achieve better performance in \techphaseA{} and \techphaseB{} when provided with a small number of examples.


\textbf{Critique (\critique{}).}
The critique prompt is designed to elicit evaluative feedback from an LLM; it asks the model to review, critique, or refine an existing generation by pointing out errors or suggesting improvements \cite{sun2025source}. 
In this study, we employ \critique{} prompts that explicitly point out errors in the LLM's outputs to specific tasks, such as the inapplicable MRs produced in \techphaseA{} and invalid meta drivers constructed in \techphaseB{}.



\textbf{Expert (\expert{}).} 
The expert prompt instructs the LLM to assume the role of a domain expert, which biases its output toward specialized knowledge and style \cite{xu2023expertprompting}.
In this study, we prepare the four other prompts with minimal expert-guided instructions, such as ``\textit{you are a software testing expert familiar with metamorphic testing}'' and ``\textit{you are an expert in fuzz testing}'', to ensure that the LLM can handle \mfoe{} tasks properly. 
In addition to these basic instructions, we provide restrictive instructions to guide the LLM to focus more on specific tasks (i.e., \techphaseA{} and \techphaseB{}).  
Specifically, we ask an LLM to \lineno{1} understand \techphaseA{} or \techphaseB{} by dissecting it into concrete steps and \lineno{2} generate a description of an expert competent for the given task from a second-person perspective. 

\textbf{Chain-of-thought (\chain{}).} 
The chain-of-thought prompt encourages the LLM to explicitly articulate intermediate reasoning steps, helping the model better understand the task and thus improve its performance \cite{wei2022chain}.
To construct appropriate \chain{} templates for \techphaseA{} and \techphaseB{}, we summarize general workflows from prior studies on metamorphic testing \cite{chen2018metamorphic} and fuzz driver generation \cite{babic2019fudge}, and instantiate them as structured operation chains.
Details of the \chain{} prompts for \techphaseA{} and \techphaseB{} are provided in \autoref{sec:promptA} and \autoref{sec:promptB}, respectively.

\subsubsection{Prompting for \techphaseA{}}
\label{sec:promptA}

\begin{figure}
    \centering
    \includegraphics[width=\linewidth]{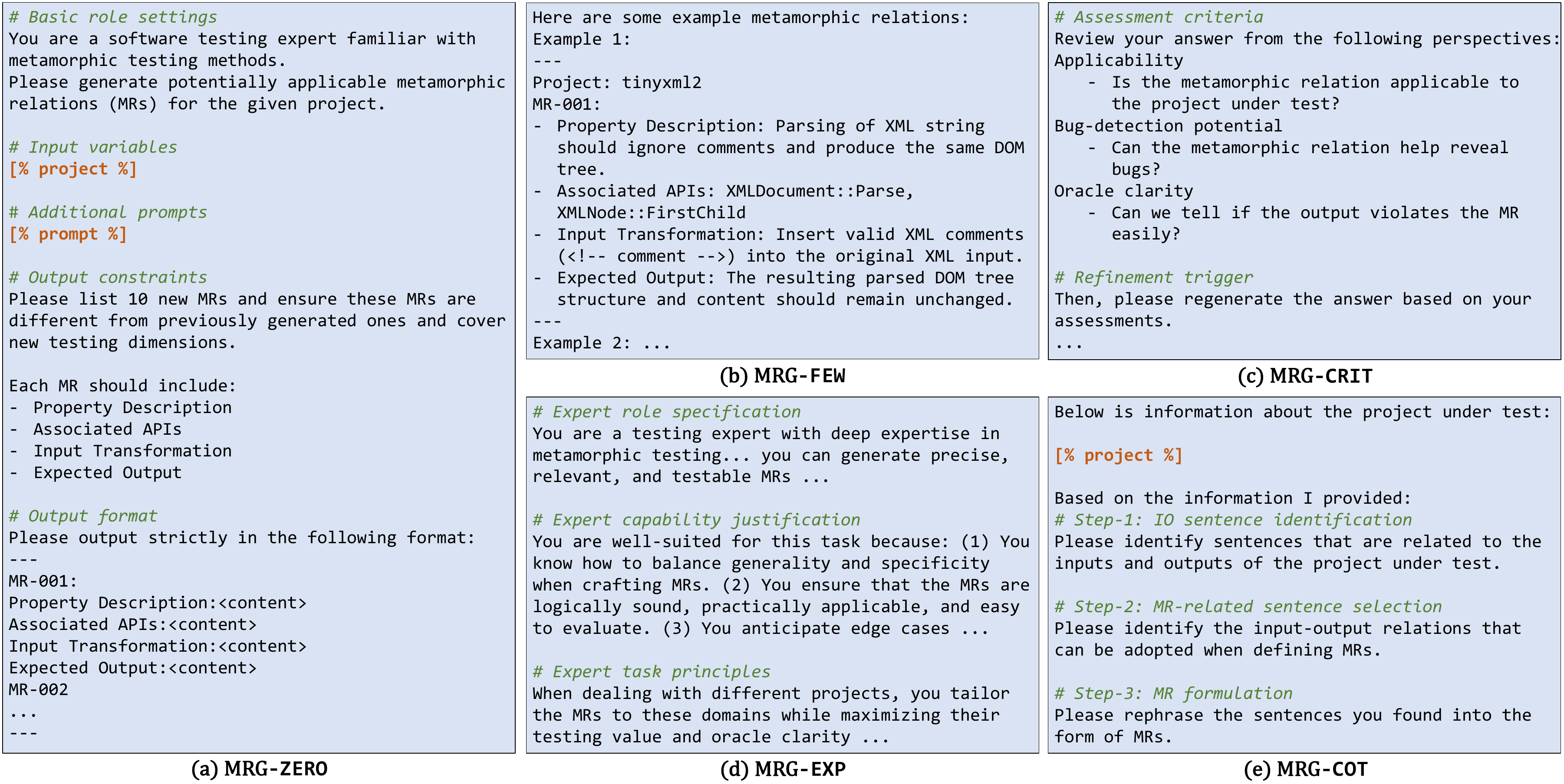}
    \caption{Prompt templates used in \techphaseA{}.}
    \label{fig:mrg-prompts}
    \Description{Prompt templates used in \techphaseA{}.}
\end{figure}

\definecolor{lightbrown}{RGB}{197, 90, 17}
\newcommand{\promptvar}[1]{\textcolor{lightbrown}{\texttt{[\% #1 \%]}}}
\autoref{fig:mrg-prompts} displays the prompt templates designed for \techphaseA{}.
For clarity, we simplify these prompts by omitting non-critical content.
The text segments formatted as \promptvar{var} are prompt variables that will be instantiated during prompt construction. 
For example, the \promptvar{project} field will be replaced with concrete information (e.g., documentation and source code) of the library under test, guiding the LLM to generate MRs for the library.

\autoref{fig:mrg-prompts}(a) presents the \zero{} template, which also serves as the prompt skeleton. 
When constructing prompts for \few{}, \critique{}, \expert{}, and \chain{}, the \promptvar{prompt} field in this skeleton is replaced with the extended content shown in \autoref{fig:mrg-prompts}(b)\rangeline{}(e).
In addition to the variables, the \zero{} prompt consists of three components:
\lineno{1} \textit{Basic role settings}, which specify the fundamental expertise and task goals of the LLM.
\lineno{2} \textit{Output constraints}, which instruct the LLM to generate 10 distinct MRs.
\lineno{3} \textit{Output formatting}, which stipulates the structure of the generated MRs.

\autoref{fig:mrg-prompts}(b) presents the additional prompt for \few{}, where two examples are provided to help the LLM understand the goal of \techphaseA{} and illustrate the expected format of the generated MRs. 
To construct these examples, we first manually analyze two real-world projects, TinyXML2 and LibPNG, and extract one representative MR from each; the extracted MRs are then organized according to the \zero{} template.
Both projects are integrated in \ossfuzz{} and are widely used in fuzzing evaluation, making them suitable examples for \mfoe{}.

\autoref{fig:mrg-prompts}(c) presents the additional prompt for \critique{}, where the LLM is instructed to review the previously generated MRs from three aspects: applicability, bug-detection potential, and oracle clarity. 
The detailed descriptions of these aspects are provided immediately afterward to facilitate the LLM’s understanding.
Through the self-critique process, we aim for the generated MRs to be not only applicable to the target library but also testable and easy to validate by humans. 

\autoref{fig:mrg-prompts}(d) presents the additional prompt for \expert{}, which comprises three parts: 
\lineno{1} \textit{Expert role specification}, which further specifies the LLM expertise beyond the basic ones provided in the \zero{} template.
\lineno{2} \textit{Expert capability justification}, which defines criteria for high-quality MRs and facilitates the reasoning process.
\lineno{3} \textit{Expert task principles}, which decomposes and constrains the \techphaseA{} task from the expert perspective. 
Following prior work \cite{xu2023expertprompting, sun2025source}, we construct the \expert{} prompt with the assistance of an LLM via few-shot prompting.
Specifically, we adopt the task description for code summarization provided by \citet{sun2025source} as an example, and supply the key steps of \techphaseA{} (e.g., understanding the provided documentation and extracting I/O relations) to prompt ChatGPT (powered by GPT-4o) to synthesize the \expert{} template.

\autoref{fig:mrg-prompts}(e) presents the additional prompt for \chain{}, where \techphaseA{} is divided into three steps.
First, we instruct the LLM to identify and focus on sentences that describe the input–output (IO) relations of the target libraries; these IO relations are essential for effective metamorphic testing, as they guide the generation of follow-up test cases and underpin the validation of MRs (described in \autoref{sec:bk-mt}). 
Second, we instruct the LLM to filter out IO relations that are \textit{not} suitable for defining MRs.
For example, in an XML parsing project such as TinyXML2, inserting comments into a document will not change its parsed structure, which constitutes a suitable IO relation to define an MR. 
In contrast, changing the order of attributes within an XML element should not change the element's semantics, rendering it seemly an appropriate IO relation for defining an MR; however, this relation may lead to different serialized outputs, which makes it difficult to validate reliably and thus unsuitable for MR construction. 
Third, for the remaining IO relations, we instruct the LLM to convert them into MRs according to the format defined in the \zero{} template.


\subsubsection{Prompting for \techphaseB{}}
\label{sec:promptB}

\begin{figure}
    \centering
    \includegraphics[width=\linewidth]{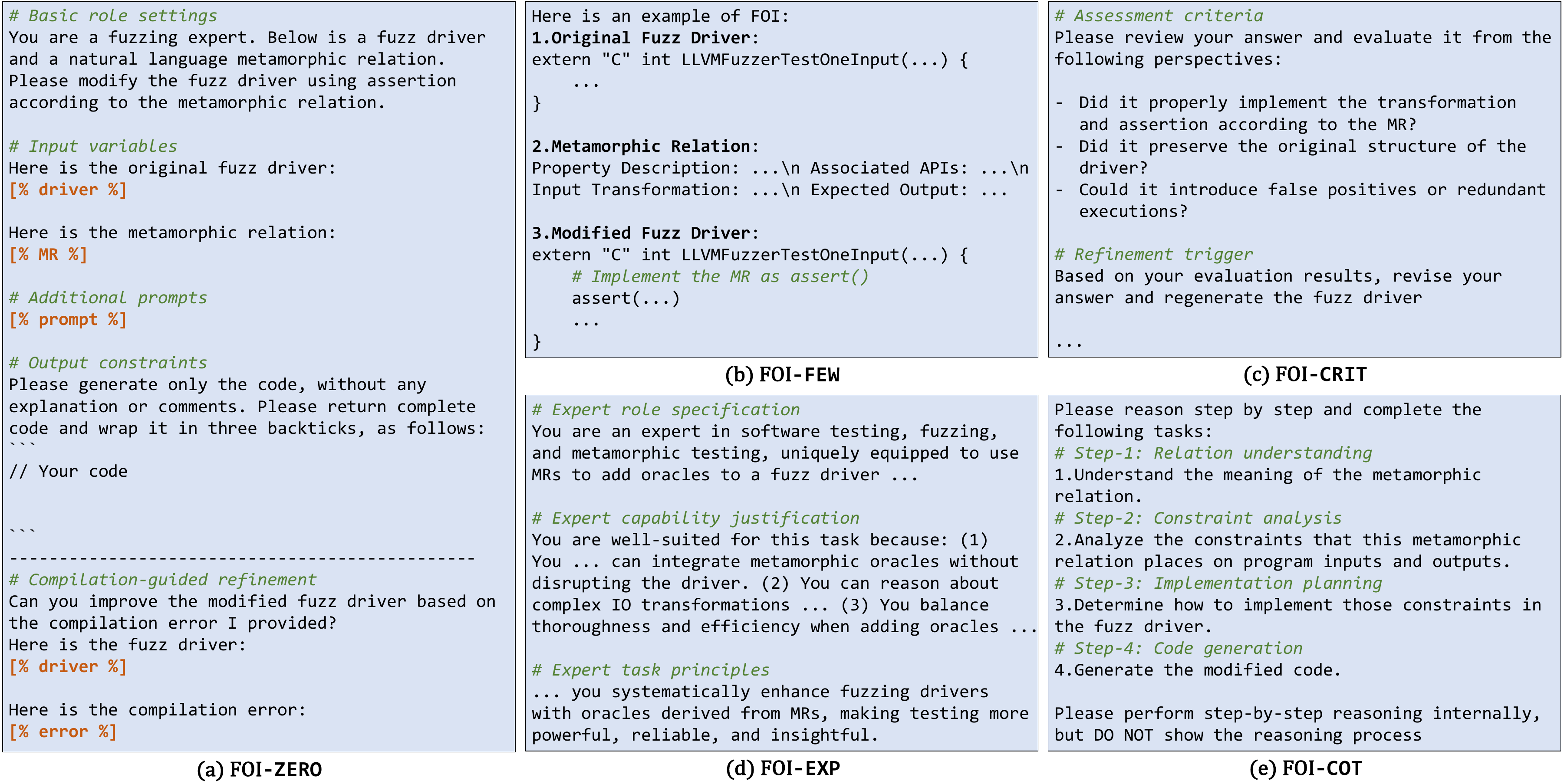}
    \caption{Prompt templates used in \techphaseB{}.}
    \label{fig:doe-prompts}
    \Description{Prompt templates used in \techphaseB{}.}
\end{figure}

\autoref{fig:doe-prompts} displays the prompt templates for \techphaseB{}.
Similarly to \autoref{fig:mrg-prompts}, \autoref{fig:doe-prompts}(a) presents the \zero{} template (the prompt skeleton), and \autoref{fig:doe-prompts}(b)\rangeline{}\autoref{fig:doe-prompts}(e) present the additional prompts for specific strategies.
As shown in \autoref{fig:doe-prompts}(a), \techphaseB{} takes two input variables: \lineno{1} the original driver to be enhanced (i.e., \promptvar{driver}) and \lineno{2} the MR to be instantiated as a fuzz oracle (i.e., \promptvar{MR}).
After setting the basic role of the LLM as a fuzzing expert and specifying the input variables, the \zero{} template further constrains the output format and instructs the LLM to output only code.
The optional prompt below the dashed line is for compilation-level driver validation (described in \autoref{subsubsec:phaseB}); it instructs the LLM to resolve the compilation issues of the meta driver candidates using the error messages as guidance. 


In \autoref{fig:doe-prompts}(b), the \few{} prompt includes an example that demonstrates how to implement and incorporate a metamorphic-based oracle into a given driver. 
The example comprises three parts: \lineno{1} the original fuzz driver, \lineno{2} the MR to be implemented, and \lineno{3} the modified fuzz driver equipped with the new metamorphic oracle, which is implemented as an \texttt{assert}. 
We craft this example by manually implementing the MR ``\textit{Adding an attribute to XML element should increase the attribute count of the element by 1}'' in a driver from the TinyXML2 project.
Details of the prompt can be found in our artifact \cite{techArtifact}.

In \autoref{fig:doe-prompts}(c), the \critique{} prompt first instructs the LLM to review its previous output to ensure that \lineno{1} the given MR is correctly implemented, \lineno{2} the modified driver remains consistent with its original structure, and \lineno{3} the modification does \textit{not} introduce false positives or unnecessary executions.
Based on the review results, the prompt further instructs the LLM to revise its output and re-generate the modified fuzz drivers if any of the above checks are violated.


In \autoref{fig:doe-prompts}(d), the \expert{} prompt further strengthens the LLM's capability to perform \techphaseB{} by \lineno{1} specifying stronger expertise in fuzzing and metamorphic testing, \lineno{2} highlighting key considerations in implementing metamorphic-based oracles for fuzzing, and \lineno{3} concretizing the criteria of accomplishing \techphaseB{} with high quality.
Consistent with \techphaseA{}, the \expert{} prompt for \techphaseB{} is likewise synthesized with the assistance of ChatGPT via few-shot prompting.


In \autoref{fig:doe-prompts}(e), the \chain{} prompt instructs the LLM to perform \techphaseB{} in four steps.
First, analyze the given MR to understand its intention.
Second, identify the constraints intended by the MR, which define how outputs change in response to input transformations.
Third, determine how to implement the identified constraints within the given driver's context and plan the modification accordingly.
Fourth, trigger the generation of the meta driver.
Collectively, these four steps are designed based on the general workflows of fuzz driver generation and metamorphic testing (see \autoref{sec:bk}). 

\begin{figure*}

    \begin{minipage}[b]{.49\linewidth}
        \centering
        \includegraphics[width=\linewidth]{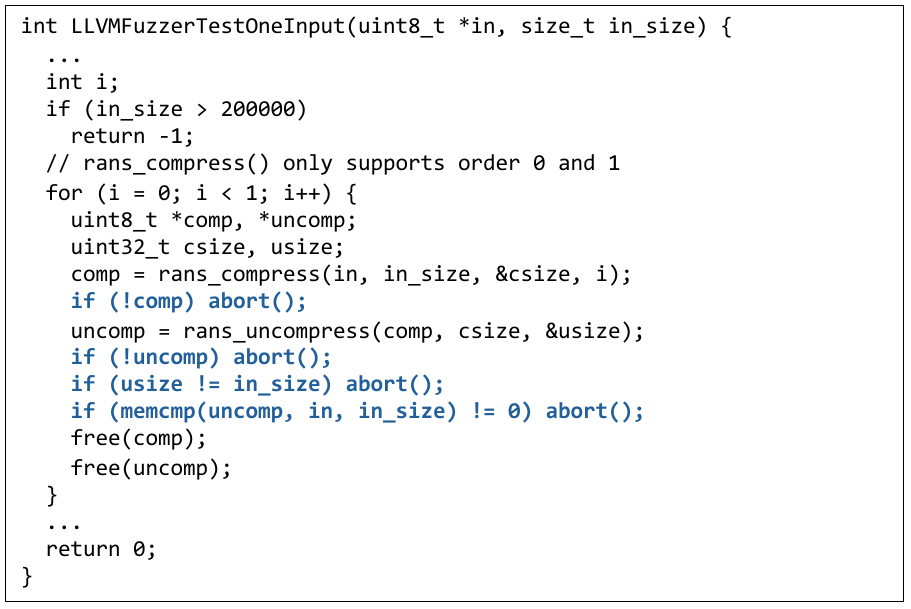}
        \subcaption{
            An Abort oracle in driver \texttt{entropy\_fuzz} from project Htslib \cite{htslibentropyfuzz}.
        }
        \label{fig:abort-oracle}
    \end{minipage}
    \begin{minipage}[b]{.49\linewidth}
        \centering
        \includegraphics[width=\linewidth]{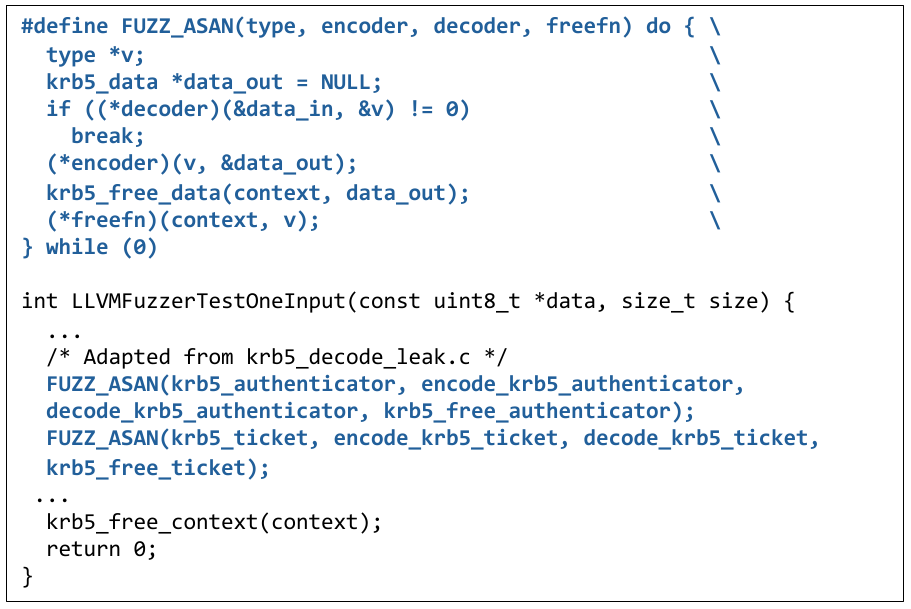}
        \subcaption{
            A Sanitizer oracle in driver \texttt{fuzz\_asn} from project Krb5 \cite{krb5fuzzasn}. 
        }
        \label{fig:sanitizer-oracle}
    \end{minipage}
    \caption{Example drivers implementing Abort and Sanitizer oracles.}
    \Description{Examples of Abort and Sanitizer Oracles}
\end{figure*}

\section{Results}
This section presents the study results, which are organized by the sequence of RQs (\autoref{subsec:rqs}). 
For each RQ, we first summarize the key findings and then present the detailed results.

\subsubsection*{Experimental Setup for \mfoe{}}
Our study comprises two group of experiments: running \techname{} (i.e., \techphaseA{} and \techphaseB{}) and fuzzing with meta drivers.
All experiments are conducted on a server equipped with an Intel(R) Xeon(R) Silver 4316 CPU @ 2.30GHz (80 logical cores) and 512 GB of RAM, running Ubuntu 20.04.6 LTS.
In total, \techname{} consumes \ntokens{} tokens.
For fuzzing, we run each driver to for \fuzzhours{} hours.
For projects that prepare seed sets in the original build scripts, we directly use these seeds as initial seeds.
For the remaining projects that do not explicitly provide initial seeds, we collect seeds by crawling the internet or from the dataset provided by \citet{herrera2021seed}.
All initial seeds are available in our artifact \cite{techArtifact}.

\subsection{\rqline{1}{Existing Fuzz Oracles}}
\label{subsec:rq-existing-oracles}

\findingbox{1}{Existing fuzz oracles can be classified into four types, namely None, Assert, Abort, and Sanitizer. 
Among the \nossfuzzdrivers{} drivers that we collect from \ossfuzz{}, the four oracle types account for 63.9\%, 18.3\%, 16.1\%, and 1.8\%, respectively.
Considering both the types and purposes, 81.1\% of the existing fuzz drivers do not implement any functional oracles to verify the APIs under test.
These observations highlight the critical need for (automatic) fuzz oracle enhancement.
}

\subsubsection{Collection of Fuzz Drivers}
\label{subsubsec:fuzz-oracles-collect}
We collect fuzz drivers from \ossfuzz{} in three steps.
First, we implement scripts to download and build \ossfuzz{} projects on a local machine.
Second, based on the build results, we identify driver source files and record driver names with another Python script.
Through these two steps, we obtain \tocheck{1,078} plausible drivers.
Third, we manually inspect drivers' source code to exclude those that appear to be drivers but are actually not (e.g., the AFL glue code \texttt{afl\_driver.c}).
As a result, we filter out 10 fake drivers and finally obtain \nossfuzzdrivers{} drivers from \nossfuzzprojects{} projects.
All drivers and scripts can be found in our artifact \cite{techArtifact}.

\subsubsection{Analysis of Existing Fuzz Oracles}
Based on the source code of the collected drivers, we classify the oracles implemented in these drivers into four types. 
The details are as follows:

\begin{enumerate}[leftmargin=*, topsep=3pt]
    \item \textbf{None Oracle}. 
    A fuzz driver is classified as implementing a None oracle if it contains no dedicated bug-detection mechanisms and relies solely on program crashes (e.g., memory corruptions) to indicate bugs.
    An example of a None oracle is displayed in \autoref{fig:withoutoracle}. 

    \item \textbf{Assert Oracle}. 
    A fuzz driver is classified as implementing an Assert oracle if it invokes standard \texttt{assert} statements or custom assertion mechanisms.
    An example driver with an Assertion oracle is shown in \autoref{fig:withoracle}, which includes two standard \texttt{assert} statements.

    \item \textbf{Abort Oracle}. 
    A fuzz driver is classified as implementing an Abort oracle if it raises abnormal exits.
    Given that most of the \ossfuzz{} drivers are LibFuzzer-style, we treat invocations of \texttt{abort}, \texttt{\_\_builtin\_trap}, and \texttt{exit} (with non-zero exit code) statements as abnormal exits, as LibFuzzer leverages these statements to signal crashing or failing inputs.
    \autoref{fig:abort-oracle} illustrates an example of an Abort oracle, where the driver invokes \texttt{abort} (colored blue) when the values of certain variables are not as expected.
    
    \item \textbf{Sanitizer Oracle}. 
    A fuzz driver implements mechanisms to facilitate sanitizers (e.g., ASAN \cite{asan} and MSAN \cite{msan}) in detecting erroneous behaviors.
    \autoref{fig:sanitizer-oracle} illustrates an example of a Sanitizer oracle, where the driver first defines a macro \texttt{FUZZ\_ASAN} to assist ASAN in detecting bugs by triggering carefully crafted memory manipulations, and then invokes this macro multiple times in the driver body.
\end{enumerate}

\begin{figure*}

    \begin{minipage}[b]{.325\linewidth}
        \centering
        \includegraphics[width=\linewidth]{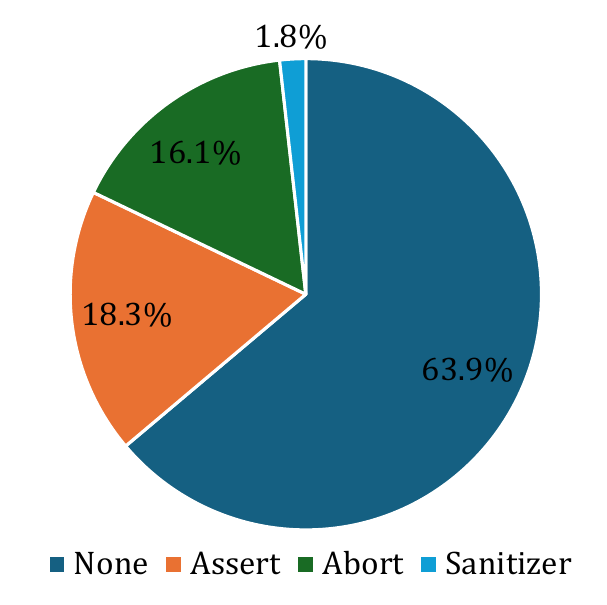}
        \subcaption{
            Distribution of None, Assert, Abort, and Sanitizer oracles. 
        }
        \label{subfig:props-oracle-types}
    \end{minipage}
    \begin{minipage}[b]{.325\linewidth}
        \centering
        \includegraphics[width=\linewidth]{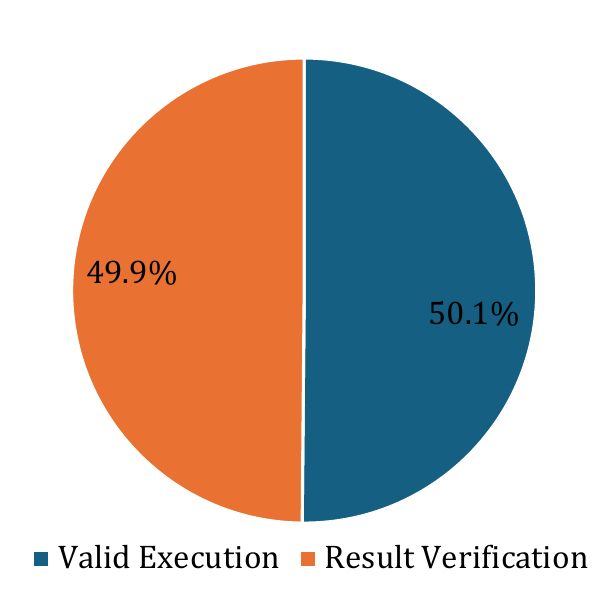}
        \subcaption{
            Distribution of purposes of the \nassertabort{} Assert and Abort oracles.
        }
        \label{subfig:props-oracle-purposes}
    \end{minipage}
    \begin{minipage}[b]{.325\linewidth}
        \centering
        \includegraphics[width=\linewidth]{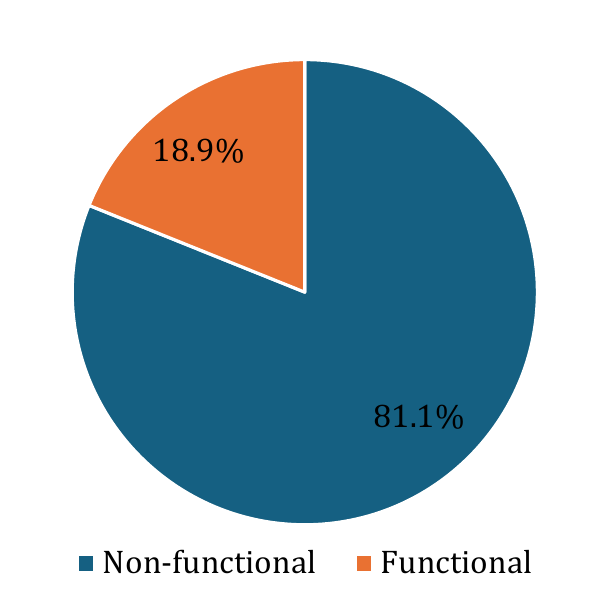}
        \subcaption{
        Distribution of functional and non-functional fuzz oracles.
        }
        \label{subfig:props-functionality}
    \end{minipage}
    \caption{Proportions of fuzz oracles the \nossfuzzdrivers{} \ossfuzz{} drivers.}
    \Description{Statistics of fuzz oracles in \nossfuzzdrivers{} \ossfuzz{} drivers.}
\end{figure*}

\autoref{subfig:props-oracle-types} shows the proportions of different types of fuzz oracles. 
Among the \nossfuzzdrivers{} \ossfuzz{} drivers that we collect, 63.9\% (682) implement \textit{no} explicit oracles (i.e., None type) and rely on crashes to indicate bugs. 
This observation aligns with common practice in modern fuzzing, where crashes are often used as implicit oracles.
In addition to the None type, 18.3\% (195) of the drivers implement Assert oracles by invoking at least one \texttt{assert} statement, and 16.3\% (172) implement Abort oracles.
The Sanitizer oracles account for 1.8\% (19), which is the lowest among the four types of oracles.

We observe that although some drivers implement Assert or Abort oracles, these oracles may not be intended to verify the tested APIs. 
Inspired by this observation, we further inspect the source code of the 367 drivers that implement Assert and Abort oracles to identify the exact purposes.
Two types of oracle purposes are observed:
\lineno{1} \textit{Result verification}, for which the oracle checks the return values of tested APIs and alarms when the values are not as expected. 
\lineno{2} \textit{Valid execution}, for which the oracle enforces certain preconditions (e.g., a previous API call succeeds and returns an expected value) to ensure that the driver is executed with valid states and inputs.

\autoref{subfig:props-oracle-purposes} visualizes the proportion of purposes among the Assert and Abort oracles, where we observe that 50.1\% (184) of the drivers implement oracles only for valid execution; this observation confirms that only a few drivers implement functional oracles. 
To quantify the proportion of functional fuzz oracles, we label Sanitizer oracles and Assert/Abort oracles used for result verification as functional, and treat the remaining ones as non-functional.
As shown in \autoref{subfig:props-oracle-purposes}, only 18.9\% of the \nossfuzzdrivers{} drivers implement functional oracles, underscoring the urgent need to improve fuzz oracles. 
Given that metamorphic-based oracles can provide strong support for functional verification of the APIs under test, these observations highlight the potential of \mfoe{} and motivate our study.

\subsection{\rqline{2}{Evaluation of LLM-based \techphaseA{}}}
\label{subsec:rq-eval-mrg}

\findingbox{2}{
\techname{} generates \tocheck{272\rangeline{}485} MRs for the \nmfoesubjects{} projects, with an overall applicable proportion of \tocheck{77.3\%}. 
These results demonstrate the feasibility of LLM-based \techphaseA{} with promising effectiveness.
The applicable proportion of generated MRs is weakly negatively correlated with the LoC of the projects, with a Pearson Correlation Coefficient of \tocheck{-0.38}. 
}

\subsubsection{Subjects of \mfoe{} Experiments}
The columns 1\rangeline{}3 of \autoref{tab:mfoe-subjects} present the projects and drivers selected for the \mfoe{} experiments.
Specifically, we experiment with \nmfoesubjects{} projects from \ossfuzz{}, and select one representative driver for each project.
These projects are widely used in fuzzing research, strengthening the validity of our study \cite{zheng2025mendelfuzz,TOSEM25-bench}.

\subsubsection{\techphaseA{} Statistics}

The columns 4\rangeline{}6 of \autoref{tab:mfoe-subjects} show the \techphaseA{} results obtained in each subject, where the table heads \#R, \#UR, and \#AR denote the counts of generated, unique, and applicable MRs; the last row presents the totals.
We report the counts for \#R and \#UR, and the counts and proportions for \#AR. 
The applicable proportion (AP) is calculated as $\frac{\text{\#AR}}{\text{\#R}}\times100\%$, which is shown in the table with smaller fonts.
As shown, \techname{} generates 272\rangeline{}485 MRs for the \nmfoesubjects{} subjects, with an overall AP of 77.3\%.
These results demonstrate the feasibility of LLM-based \techphaseA{}, with a reasonably high AP exceeding 70\%.
AP varies across subjects (67.8\%–90.0\%), suggesting that the generation of applicable MRs depends on the characteristics of the project under test.
To further investigate this dependency, we compute the Pearson correlation coefficient (PCC) \cite{cohen2009pearson} between AP and the lines of code (LoC) of the projects (reported in \autoref{tab:LoC-calls-crashes}).
The PCC is -0.38, indicating a weak negative correlation between AP and LoC.
This result is somewhat counterintuitive, as larger projects are expected to contain more constraints and thus should  yield more MRs.

\begin{table}
\small
    \centering
    \caption{Experimental subjects and \mfoe{} results for each project.}
    \label{tab:mfoe-subjects}
\begin{tabular}{|c|c|c|r|r|r|r|r|r|}
\hline
\textbf{ID} & \multicolumn{1}{l|}{\textbf{Project}} & \multicolumn{1}{l|}{\textbf{Driver}} & \textbf{\#R} & \textbf{\#UR} & \textbf{\#AR} & \textbf{\#D} & \textbf{\#CD} & \textbf{\#VD} \\
\hline
\textbf{S01} & \multicolumn{1}{l|}{Rapidjson} & \multicolumn{1}{l|}{\texttt{fuzzer}} & 300   & 285   & 270 \tiny{(90.0\%)} & 1245  & 1198  & 779 \tiny{(62.6\%)} \\
\hline
\textbf{S02} & \multicolumn{1}{l|}{Tinyxml2} & \multicolumn{1}{l|}{\texttt{driver}} & 326   & 313   & 286 \tiny{(87.7\%)} & 1290  & 1254  & 614 \tiny{(47.6\%)} \\
\hline
\textbf{S03} & \multicolumn{1}{l|}{Brotli} & \multicolumn{1}{l|}{\texttt{decode\_fuzzer}} & 406   & 385   & 309 \tiny{(76.1\%)} & 1440  & 1367  & 811 \tiny{(56.3\%)} \\
\hline
\textbf{S04} & \multicolumn{1}{l|}{Libpng} & \multicolumn{1}{l|}{\texttt{libpng\_read\_fuzzer}} & 294   & 272   & 230 \tiny{(78.2\%)} & 1076  & 662   & 389 \tiny{(36.2\%)} \\
\hline
\textbf{S05} & \multicolumn{1}{l|}{Zlib} & \multicolumn{1}{l|}{\texttt{zlib\_uncompress\_fuzzer}} & 319   & 291   & 244 \tiny{(76.5\%)} & 1160  & 1143  & 906 \tiny{(78.1\%)} \\
\hline
\textbf{S06} & \multicolumn{1}{l|}{Libyaml} & \multicolumn{1}{l|}{\texttt{libyaml\_parser\_fuzzer}} & 290   & 282   & 225 \tiny{(77.6\%)} & 1015  & 815   & 265 \tiny{(26.1\%)} \\
\hline
\textbf{S07} & \multicolumn{1}{l|}{Liblouis} & \multicolumn{1}{l|}{\texttt{fuzz\_backtranslate}} & 272   & 268   & 201 \tiny{(73.9\%)} & 830   & 680   & 337 \tiny{(40.6\%)} \\
\hline
\textbf{S08} & \multicolumn{1}{l|}{C-ares} & \multicolumn{1}{l|}{\texttt{ares-test-fuzz}} & 485   & 451   & 369 \tiny{(76.1\%)} & 1680  & 1086  & 782 \tiny{(46.5\%)} \\
\hline
\textbf{S09} & \multicolumn{1}{l|}{Re2} & \multicolumn{1}{l|}{\texttt{re2\_fuzzer}} & 407   & 398   & 276 \tiny{(67.8\%)} & 1315  & 1210  & 732 \tiny{(55.7\%)} \\
\hline
\textbf{S10} & \multicolumn{1}{l|}{Libtiff} & \multicolumn{1}{l|}{\texttt{tiff\_read\_rgba\_fuzzer}} & 376   & 373   & 275 \tiny{(73.1\%)} & 1300  & 1192  & 613 \tiny{(47.2\%)} \\
\hline
\multicolumn{3}{|c|}{\textbf{Total}} & 3475  & 3318  & 2685 \tiny{(77.3\%)} & 12351 & 10607 & 6228 \tiny{(50.4\%)} \\
\hline
\end{tabular}%
\end{table}

\subsection{\rqline{3}{Evaluation of LLM-based \techphaseB{}}}
\label{subsec:rq-driver-enhance}

\findingbox{3}{Based on the generated MRs, \techname{} implements \tocheck{830\rangeline{}1,680} meta driver candidates for the \nmfoesubjects{} subjects, among which \tocheck{6,228 (50.4\%)} are valid in total.
Extensive fuzzing with groups of \fuzzhours{}-hour campaigns show that: 
\lineno{1} On average, the meta drivers cover \tocheck{18.7\%} more edges than the original drivers by encompassing \tocheck{38.0} more LoC and \tocheck{4.1} more API calls.
\lineno{2} \tocheck{30\rangeline{}508} unique crashes are triggered by the meta drivers on the \nmfoesubjects{} projects, \tocheck{eight} of which the original drivers fail to detect any crashes. 
These results indicate that metamorphic-based oracles can improve both edge coverage and crash discovery.
}

\subsubsection{\techphaseB{} Statistics}
The columns 7\rangeline{}9 of \autoref{tab:mfoe-subjects} exhibit the \techphaseB{} results, where \#D, \#CD, and \#VD denote the numbers of implemented, compilable, and valid drivers.
Similarly, we present the valid proportion (VP) in the tiny font, which is calculated as $\frac{\text{\#VD}}{\text{\#D}}\times100\%$.
Based on applicable MRs, \techname{} implements metamorphic-based oracles and crafts \tocheck{830\rangeline{}1,680} meta drivers.
The VPs of the implemented meta drivers range from \tocheck{26.1\% to 78.1\%} among subjects, with the lowest VP observed in Libyaml and the highest in Zlib.
The overall VP is \tocheck{50.4\%}, indicating that about half of the meta drivers are invalid. 
Losses of meta driver candidates are observed in both validation stages of \techphaseB{} (described in \autoref{subsubsec:phaseB}), demonstrating the necessity of the two-stage design.
Specifically, \tocheck{14.1\%} of the candidates are discarded by compilation-level validation, indicating that generating compilable meta drivers remains a challenge for LLMs. 
A much larger proportion (\tocheck{85.9\%}) of candidates are discarded by runtime-level validation, suggesting that LLMs have limited ability to predict runtime behaviors of generated drivers.
Among those discarded during runtime-level validation, \tocheck{98.7\%} are removed because they trigger excessive crashes, while the rest are discarded for achieving too little coverage.
These results indicate that LLMs tend to implement metamorphic-based oracles that induce false positives, reflecting the limitations of modern LLMs in \techphaseB{}.



\begin{table}
\small
  \centering
  \caption{LoC, API calls, and crashes found in each project.}
\begin{tabular}{|c|r|r|r|r|r|r|r|r|r|}
\hline
      & \multicolumn{3}{c|}{\textbf{LoC}} & \multicolumn{2}{c|}{\textbf{API Calls}} & \multicolumn{2}{c|}{\textbf{Unique MC}} & \multicolumn{2}{c|}{\textbf{Unique AF}} \\
\hline
\textbf{ID} & \textbf{Project} & \textbf{Orig. D} & \textbf{Avg. VMD} & \textbf{Orig. D} & \textbf{Avg. VMD} & \textbf{Orig. D} & \textbf{VMD} & \textbf{Orig. D} & \textbf{VMD} \\
\hline
\textbf{S01} & 29514 & 36    & 68.0  & 10    & 18.7  & 0     & 4     & 1     & 504 \\
\hline
\textbf{S02} & 5820  & 20    & 61.2  & 7     & 17.1  & 0     & 3     & 0     & 162 \\
\hline
\textbf{S03} & 38819 & 47    & 78.5  & 3     & 5.2   & 0     & 2     & 0     & 28 \\
\hline
\textbf{S04} & 58206 & 156   & 203.8 & 39    & 43.7  & 0     & 6     & 0     & 38 \\
\hline
\textbf{S05} & 30337 & 13    & 41.0  & 2     & 5.3   & 0     & 6     & 0     & 102 \\
\hline
\textbf{S06} & 10601 & 26    & 82.7  & 5     & 8.2   & 0     & 9     & 0     & 99 \\
\hline
\textbf{S07} & 37447 & 80    & 112.9 & 20    & 20.8  & 1     & 8     & 0     & 121 \\
\hline
\textbf{S08} & 44750 & 239   & 270.9 & 55    & 54.6  & 0     & 6     & 0     & 34 \\
\hline
\textbf{S09} & 27921 & 198   & 227.0 & 56    & 59.9  & 0     & 6     & 0     & 326 \\
\hline
\textbf{S10} & 94402 & 188   & 236.9 & 31    & 35.1  & 0     & 12    & 0     & 52 \\
\hline
\end{tabular}%
  \label{tab:LoC-calls-crashes}%
\end{table}%

\subsubsection{Code Coverage}

The upper subtable of \autoref{tab:fuzz-caps-means} reports the average edge coverage achieved in \fuzzhours{} hours of fuzzing.
The Orig column reports the edges covered by the original driver, averaged over five runs, while the All column reports the averages across all valid meta drivers.
As shown, the valid meta drivers cover more edges (colored gray) in seven of the subjects, with increases of \tocheck{6.2\%\rangeline{}81.9\%}.
Decreases are also observed on the remaining subjects Libpng and Libtiff, with slight decreases of \tocheck{1.0\%} and 1.2\%. 
On average, the valid meta drivers cover \tocheck{18.7\%} more edges than the original drivers, a significant improvement beyond our expectations, since \techname{} is primarily designed to improve oracles (bug detection) rather than increase code coverage.

To understand the changes in coverage, we measure \lineno{1} the lines of code and \lineno{2} the number of API calls in the original and the valid meta drivers.
The results are exhibited in \autoref{tab:LoC-calls-crashes}, where Orig. D denotes statistics of the original driver and Avg. VMD denotes averages of the valid meta drivers.
Compared to the original driver, the meta drivers contain  \tocheck{28.0\rangeline{}56.7} more LoC, with an average increase of 38.0 LoC.
For API calls, meta drivers invoke \tocheck{0.8\rangeline{}10.1} more distinct APIs than the original driver on \tocheck{nine} subjects, except for S09 (Re2), where the meta drivers instead invoke 0.4 fewer APIs.
This observation indicates that \techname{} can produce meta drivers with fewer LoC than the original driver, inspiring further inspection that reveals this phenomenon in \tocheck{203} valid meta drivers. 
All these results indicate that implementing metamorphic-based oracles is dependent not only on the original API under test but also on other APIs relevant to MR verification.
 
\begin{table}
\small
  \centering
  \caption{Average covered edges and triggered crashes after \fuzzhours{} hours of fuzzing.}
\begin{tabular}{|c|r|r|r|r|r|r|r|r|r|r|}
\hline
\multicolumn{11}{|c|}{\textbf{Average \#Edges}} \\
\hline
\textbf{ID} & \textbf{Orig} & \textbf{ALL} & \textbf{GLM} & \textbf{DeepSeek} & \textbf{GPT} & \textbf{ZERO} & \textbf{FEW} & \textbf{CRIT} & \textbf{COT} & \textbf{EXP} \\
\hline
\textbf{S01} & 1627.1  & \cellcolor[rgb]{0.910, 0.910, 0.910} 1728.5  & \cellcolor[rgb]{0.753, 0.902, 0.961} 1480.3  & \cellcolor[rgb]{0.267, 0.702, 0.882} 1802.8  & \cellcolor[rgb]{0.271, 0.706, 0.886} 1801.8  & \cellcolor[rgb]{0.588, 0.827, 0.655} 1756.4  & \cellcolor[rgb]{0.988, 0.988, 1.000} 1561.1  & \cellcolor[rgb]{0.561, 0.816, 0.631} 1768.7  & \cellcolor[rgb]{0.702, 0.875, 0.753} 1700.3  & \cellcolor[rgb]{0.388, 0.745, 0.482} 1851.6  \\
\hline
\textbf{S02} & 422.1  & \cellcolor[rgb]{0.910, 0.910, 0.910} 530.9  & \cellcolor[rgb]{0.753, 0.902, 0.961} 492.7  & \cellcolor[rgb]{0.624, 0.851, 0.941} 515.6  & \cellcolor[rgb]{0.267, 0.702, 0.882} 578.0  & \cellcolor[rgb]{0.749, 0.894, 0.796} 521.8  & \cellcolor[rgb]{0.612, 0.839, 0.678} 538.5  & \cellcolor[rgb]{0.388, 0.745, 0.482} 565.5  & \cellcolor[rgb]{0.988, 0.988, 1.000} 492.6  & \cellcolor[rgb]{0.643, 0.851, 0.702} 534.9  \\
\hline
\textbf{S03} & 941.3  & \cellcolor[rgb]{0.910, 0.910, 0.910} 1712.6  & \cellcolor[rgb]{0.753, 0.902, 0.961} 1457.6  & \cellcolor[rgb]{0.576, 0.831, 0.933} 1639.4  & \cellcolor[rgb]{0.267, 0.702, 0.882} 1955.2  & \cellcolor[rgb]{0.518, 0.800, 0.596} 1719.7  & \cellcolor[rgb]{0.698, 0.871, 0.749} 1706.0  & \cellcolor[rgb]{0.514, 0.796, 0.592} 1720.0  & \cellcolor[rgb]{0.388, 0.745, 0.482} 1729.4  & \cellcolor[rgb]{0.988, 0.988, 1.000} 1683.9  \\
\hline
\textbf{S04} & \cellcolor[rgb]{0.910, 0.910, 0.910} 1722.0  & 1705.3  & \cellcolor[rgb]{0.753, 0.902, 0.961} 1648.1  & \cellcolor[rgb]{0.580, 0.831, 0.933} 1688.7  & \cellcolor[rgb]{0.267, 0.702, 0.882} 1761.3  & \cellcolor[rgb]{0.647, 0.851, 0.706} 1711.3  & \cellcolor[rgb]{0.859, 0.937, 0.890} 1698.7  & \cellcolor[rgb]{0.863, 0.937, 0.890} 1698.6  & \cellcolor[rgb]{0.988, 0.988, 1.000} 1691.0  & \cellcolor[rgb]{0.388, 0.745, 0.482} 1726.7  \\
\hline
\textbf{S05} & 327.7  & \cellcolor[rgb]{0.910, 0.910, 0.910} 486.2  & \cellcolor[rgb]{0.694, 0.878, 0.953} 464.5  & \cellcolor[rgb]{0.753, 0.902, 0.961} 452.9  & \cellcolor[rgb]{0.267, 0.702, 0.882} 547.0  & \cellcolor[rgb]{0.800, 0.914, 0.835} 483.6  & \cellcolor[rgb]{0.467, 0.780, 0.553} 490.7  & \cellcolor[rgb]{0.388, 0.745, 0.482} 492.3  & \cellcolor[rgb]{0.753, 0.894, 0.800} 484.6  & \cellcolor[rgb]{0.988, 0.988, 1.000} 479.5  \\
\hline
\textbf{S06} & 1724.1  & \cellcolor[rgb]{0.910, 0.910, 0.910} 1724.8  & \cellcolor[rgb]{0.722, 0.890, 0.957} 1710.8  & \cellcolor[rgb]{0.753, 0.902, 0.961} 1704.2  & \cellcolor[rgb]{0.267, 0.702, 0.882} 1797.6  & \cellcolor[rgb]{0.573, 0.820, 0.643} 1734.1  & \cellcolor[rgb]{0.839, 0.929, 0.871} 1706.6  & \cellcolor[rgb]{0.439, 0.765, 0.525} 1748.0  & \cellcolor[rgb]{0.988, 0.988, 1.000} 1690.8  & \cellcolor[rgb]{0.388, 0.745, 0.482} 1753.1  \\
\hline
\textbf{S07} & 1689.3  & \cellcolor[rgb]{0.910, 0.910, 0.910} 2009.5  & \cellcolor[rgb]{0.753, 0.902, 0.961} 1854.7  & \cellcolor[rgb]{0.706, 0.882, 0.953} 1888.4  & \cellcolor[rgb]{0.267, 0.702, 0.882} 2188.3  & \cellcolor[rgb]{0.663, 0.859, 0.722} 2017.4  & \cellcolor[rgb]{0.467, 0.776, 0.549} 2079.0  & \cellcolor[rgb]{0.773, 0.902, 0.816} 1983.2  & \cellcolor[rgb]{0.988, 0.988, 1.000} 1915.8  & \cellcolor[rgb]{0.388, 0.745, 0.482} 2102.4  \\
\hline
\textbf{S08} & 1404.3  & \cellcolor[rgb]{0.910, 0.910, 0.910} 1513.9  & \cellcolor[rgb]{0.753, 0.902, 0.961} 1383.8  & \cellcolor[rgb]{0.525, 0.808, 0.925} 1498.7  & \cellcolor[rgb]{0.267, 0.702, 0.882} 1625.3  & \cellcolor[rgb]{0.471, 0.780, 0.553} 1559.4  & \cellcolor[rgb]{0.388, 0.745, 0.482} 1574.5  & \cellcolor[rgb]{0.937, 0.969, 0.957} 1469.5  & \cellcolor[rgb]{0.988, 0.988, 1.000} 1459.2  & \cellcolor[rgb]{0.682, 0.867, 0.737} 1518.2  \\
\hline
\textbf{S09} & 5502.8  & \cellcolor[rgb]{0.910, 0.910, 0.910} 5522.3  & \cellcolor[rgb]{0.737, 0.898, 0.961} 5502.6  & \cellcolor[rgb]{0.753, 0.902, 0.961} 5500.7  & \cellcolor[rgb]{0.267, 0.702, 0.882} 5559.6  & \cellcolor[rgb]{0.988, 0.988, 1.000} 5499.4  & \cellcolor[rgb]{0.831, 0.925, 0.863} 5510.2  & \cellcolor[rgb]{0.663, 0.855, 0.718} 5521.6  & \cellcolor[rgb]{0.416, 0.757, 0.506} 5538.1  & \cellcolor[rgb]{0.388, 0.745, 0.482} 5539.8  \\
\hline
\textbf{S10} & \cellcolor[rgb]{0.910, 0.910, 0.910} 5530.7  & 5464.8  & \cellcolor[rgb]{0.753, 0.902, 0.961} 5354.5  & \cellcolor[rgb]{0.529, 0.812, 0.925} 5470.1  & \cellcolor[rgb]{0.267, 0.702, 0.882} 5604.3  & \cellcolor[rgb]{0.537, 0.808, 0.612} 5488.2  & \cellcolor[rgb]{0.553, 0.812, 0.624} 5484.0  & \cellcolor[rgb]{0.502, 0.792, 0.580} 5498.1  & \cellcolor[rgb]{0.988, 0.988, 1.000} 5358.4  & \cellcolor[rgb]{0.388, 0.745, 0.482} 5530.4  \\
\hline
\multicolumn{11}{|c|}{\textbf{Average \#Crashes}} \\
\hline
\textbf{ID} & \textbf{Orig} & \textbf{ALL} & \textbf{GLM} & \textbf{DeepSeek} & \textbf{GPT} & \textbf{ZERO} & \textbf{FEW} & \textbf{CRIT} & \textbf{COT} & \textbf{EXP} \\
\hline
\textbf{S01} & 2.4   & \cellcolor[rgb]{0.910, 0.910, 0.910} 27.5  & \cellcolor[rgb]{0.314, 0.722, 0.890} 30.7  & \cellcolor[rgb]{0.267, 0.702, 0.882} 31.6  & \cellcolor[rgb]{0.753, 0.902, 0.961} 21.6  & \cellcolor[rgb]{0.659, 0.855, 0.718} 27.5  & \cellcolor[rgb]{0.671, 0.859, 0.725} 27.4  & \cellcolor[rgb]{0.569, 0.820, 0.639} 28.5  & \cellcolor[rgb]{0.988, 0.988, 1.000} 24.0  & \cellcolor[rgb]{0.388, 0.745, 0.482} 30.3  \\
\hline
\textbf{S02} & 0.0   & \cellcolor[rgb]{0.910, 0.910, 0.910} 6.7  & \cellcolor[rgb]{0.753, 0.902, 0.961} 0.3  & \cellcolor[rgb]{0.267, 0.702, 0.882} 9.3  & \cellcolor[rgb]{0.345, 0.737, 0.898} 7.9  & \cellcolor[rgb]{0.388, 0.745, 0.482} 9.1  & \cellcolor[rgb]{0.725, 0.882, 0.776} 7.0  & \cellcolor[rgb]{0.867, 0.941, 0.894} 6.2  & \cellcolor[rgb]{0.988, 0.988, 1.000} 5.4  & \cellcolor[rgb]{0.961, 0.976, 0.976} 5.6  \\
\hline
\textbf{S03} & 0.0   & \cellcolor[rgb]{0.910, 0.910, 0.910} 2.1  & \cellcolor[rgb]{0.753, 0.902, 0.961} 0.9  & \cellcolor[rgb]{0.267, 0.702, 0.882} 3.3  & \cellcolor[rgb]{0.718, 0.890, 0.957} 1.1  & \cellcolor[rgb]{0.620, 0.839, 0.682} 2.7  & \cellcolor[rgb]{0.388, 0.745, 0.482} 3.8  & \cellcolor[rgb]{0.827, 0.925, 0.863} 1.7  & \cellcolor[rgb]{0.863, 0.937, 0.890} 1.5  & \cellcolor[rgb]{0.988, 0.988, 1.000} 0.9  \\
\hline
\textbf{S04} & 0.0   & \cellcolor[rgb]{0.910, 0.910, 0.910} 8.7  & \cellcolor[rgb]{0.753, 0.902, 0.961} 0.7  & \cellcolor[rgb]{0.267, 0.702, 0.882} 13.2  & \cellcolor[rgb]{0.384, 0.753, 0.902} 10.2  & 10.2  & 8.4   & \cellcolor[rgb]{0.388, 0.745, 0.482} 11.9  & \cellcolor[rgb]{0.988, 0.988, 1.000} 2.6  & \cellcolor[rgb]{0.431, 0.765, 0.522} 11.2  \\
\hline
\textbf{S05} & 0.0   & \cellcolor[rgb]{0.910, 0.910, 0.910} 2.3  & \cellcolor[rgb]{0.753, 0.902, 0.961} 1.5  & \cellcolor[rgb]{0.267, 0.702, 0.882} 2.7  & \cellcolor[rgb]{0.369, 0.745, 0.902} 2.4  & \cellcolor[rgb]{0.863, 0.937, 0.890} 2.0  & \cellcolor[rgb]{0.875, 0.945, 0.902} 2.0  & \cellcolor[rgb]{0.757, 0.898, 0.804} 2.4  & \cellcolor[rgb]{0.988, 0.988, 1.000} 1.6  & \cellcolor[rgb]{0.388, 0.745, 0.482} 3.6  \\
\hline
\textbf{S06} & 0.0   & \cellcolor[rgb]{0.910, 0.910, 0.910} 70.6  & \cellcolor[rgb]{0.753, 0.902, 0.961} 37.2  & \cellcolor[rgb]{0.506, 0.804, 0.922} 74.9  & \cellcolor[rgb]{0.267, 0.702, 0.882} 111.3  & \cellcolor[rgb]{0.988, 0.988, 1.000} 48.3  & \cellcolor[rgb]{0.388, 0.745, 0.482} 112.9  & \cellcolor[rgb]{0.725, 0.882, 0.776} 76.7  & \cellcolor[rgb]{0.835, 0.925, 0.867} 65.1  & \cellcolor[rgb]{0.933, 0.969, 0.953} 54.2  \\
\hline
\textbf{S07} & 24.6  & \cellcolor[rgb]{0.910, 0.910, 0.910} 46.3  & \cellcolor[rgb]{0.267, 0.702, 0.882} 66.1  & \cellcolor[rgb]{0.737, 0.898, 0.961} 40.8  & \cellcolor[rgb]{0.753, 0.902, 0.961} 39.9  & \cellcolor[rgb]{0.639, 0.847, 0.698} 45.5  & \cellcolor[rgb]{0.506, 0.792, 0.584} 50.5  & \cellcolor[rgb]{0.988, 0.988, 1.000} 32.2  & \cellcolor[rgb]{0.533, 0.804, 0.608} 49.5  & \cellcolor[rgb]{0.388, 0.745, 0.482} 54.9  \\
\hline
\textbf{S08} & 0.0   & \cellcolor[rgb]{0.910, 0.910, 0.910} 1.3  & \cellcolor[rgb]{0.267, 0.702, 0.882} 2.1  & \cellcolor[rgb]{0.753, 0.902, 0.961} 0.8  & \cellcolor[rgb]{0.467, 0.784, 0.918} 1.6  & \cellcolor[rgb]{0.820, 0.922, 0.855} 1.2  & \cellcolor[rgb]{0.525, 0.800, 0.600} 1.5  & \cellcolor[rgb]{0.855, 0.937, 0.886} 1.2  & \cellcolor[rgb]{0.988, 0.988, 1.000} 1.0  & \cellcolor[rgb]{0.388, 0.745, 0.482} 1.6  \\
\hline
\textbf{S09} & 0.0   & \cellcolor[rgb]{0.910, 0.910, 0.910} 51.2  & \cellcolor[rgb]{0.475, 0.788, 0.918} 54.7  & \cellcolor[rgb]{0.753, 0.902, 0.961} 36.4  & \cellcolor[rgb]{0.267, 0.702, 0.882} 68.4  & \cellcolor[rgb]{0.714, 0.878, 0.765} 55.1  & \cellcolor[rgb]{0.925, 0.965, 0.945} 46.0  & \cellcolor[rgb]{0.953, 0.976, 0.969} 44.9  & \cellcolor[rgb]{0.988, 0.988, 1.000} 43.3  & \cellcolor[rgb]{0.388, 0.745, 0.482} 68.9  \\
\hline
\textbf{S10} & 0.0   & \cellcolor[rgb]{0.910, 0.910, 0.910} 10.5  & \cellcolor[rgb]{0.753, 0.902, 0.961} 4.2  & \cellcolor[rgb]{0.396, 0.757, 0.906} 12.6  & \cellcolor[rgb]{0.267, 0.702, 0.882} 15.7  & \cellcolor[rgb]{0.949, 0.973, 0.969} 7.7  & \cellcolor[rgb]{0.898, 0.953, 0.922} 8.6  & \cellcolor[rgb]{0.988, 0.988, 1.000} 7.1  & \cellcolor[rgb]{0.678, 0.863, 0.733} 12.0  & \cellcolor[rgb]{0.388, 0.745, 0.482} 16.7  \\
\hline
\end{tabular}%
  \label{tab:fuzz-caps-means}%
\end{table}%

\subsubsection{Crash Exposure}
The lower subtable of \autoref{tab:fuzz-caps-means} reports the average number of crashes triggered by the original and all valid meta drivers.
As shown, only the original drivers of S01 and S07 expose crashes during \fuzzhours{}-hour fuzzing, whereas the meta drivers find crashes in all projects.
Specifically, meta drivers trigger \tocheck{1.3\rangeline{}70.6} crashes on the \nmfoesubjects{} subecjts, achieving \tocheck{11.9$\times$ and 1.9$\times$} of crashes than the original drivers in S01 and S07, respectively.

We further analyze the crashes to remove duplicates and identify their types.
To this end, we tag the first line of the crash stack as a summary for each crash and deduplicate crashes based on these summaries.
Two types of crashes during this process:
\lineno{1} \textbf{Memory Corruption (MC)}.
We follow the default settings of \ossfuzz{} and compile the drivers with sanitizers (i.e., ASAN, MSAN, and UBSAN), which are designed to cause crashes when suspicious memory operations are detected.
\lineno{2} \textbf{Assertion Failure (AF)}.
We are particularly interested in this type of crash because \techname{} is devised to implement metamorphic-based oracles using \texttt{assert}.

The results of the crash analysis are shown in the rightmost four columns of \autoref{tab:LoC-calls-crashes}.
The original drivers only triggered \tocheck{two} crashes in total, where \tocheck{one MC}  from S07 and \tocheck{one AF} from S01.
For meta drivers, \tocheck{30\rangeline{}508} unique crashes are detected during \fuzzhours{}-hour fuzzing, of which \tocheck{62} are MCs and \tocheck{1,466} AFs.
These results suggest that metamorphic-based oracles can significantly improve crash detection, especially for AFs, as \techname{} implements such oracles using \texttt{assert}.
However, it is worth noting that meta drivers may introduce false positives and require additional diagnosis and triage, as also observed in prior metamorphic testing approaches \cite{su2021fully,le2014compiler}.
We further discuss false positive identification and mitigation in \autoref{subsec:case-study}.

\subsection{\rqline{4}{LLM Sensitivity}}
\label{subsec:rq-llms} 

\begin{table}
\small
  \centering
  \caption{\mfoe{} results across LLMs and prompt strategies.}
  \label{tab:results-by-llms-promts}%
\resizebox{\textwidth}{!}{
\begin{tabular}{|l|r|r|r|r|r|r|r|r|r|r|r|r|}
\hline
      & \multicolumn{3}{c|}{\textbf{GLM-4-Plus}} & \multicolumn{3}{c|}{\textbf{DeepSeek-V3.2}} & \multicolumn{3}{c|}{\textbf{GPT-4.1}} & \multicolumn{3}{c|}{\textbf{Total}} \\
\hline
\textbf{MR} & \textbf{\#} & \textbf{\#U} & \textbf{\#A} & \textbf{\#} & \textbf{\#U} & \textbf{\#A} & \textbf{\#} & \textbf{\#U} & \textbf{\#A} & \textbf{\#} & \textbf{\#U} & \textbf{\#A} \\
\hline
\textbf{\zero{}} & 200   & 187   & 138 \tiny{(69.0\%)} & 196   & 194   & 156 \tiny{(79.6\%)} & 200   & 195   & 159 \tiny{(79.5\%)} & 596   & 576   & 453 \tiny{(76.0\%)} \\
\hline
\textbf{\few{}} & 200   & 170   & 135 \tiny{(67.5\%)} & 198   & 196   & 176 \tiny{(88.9\%)} & 200   & 195   & 153 \tiny{(76.5\%)} & 598   & 561   & 464 \tiny{(77.6\%)} \\
\hline
\textbf{\critique{}} & 200   & 189   & 132 \tiny{(66.0\%)} & 194   & 191   & 167 \tiny{(86.1\%)} & 199   & 194   & 158 \tiny{(79.4\%)} & 593   & 574   & 457 \tiny{(77.1\%)} \\
\hline
\textbf{\expert{}} & 199   & 179   & 142 \tiny{(71.4\%)} & 193   & 192   & 164 \tiny{(85.0\%)} & 200   & 192   & 149 \tiny{(74.5\%)} & 592   & 563   & 455 \tiny{(76.9\%)} \\
\hline
\textbf{\chain{}} & 350   & 345   & 270 \tiny{(77.1\%)} & 465   & 448   & 370 \tiny{(79.6\%)} & 281   & 251   & 216 \tiny{(76.9\%)} & 1096  & 1044  & 856 \tiny{(78.1\%)} \\
\hline
\textbf{Total} & 1149  & 1070  & 817 \tiny{(71.1\%)} & 1246  & 1221  & 1033 \tiny{(82.9\%)} & 1080  & 1027  & 835 \tiny{(77.3\%)} & 3475  & 3318  & 2685 \tiny{(77.3\%)} \\
\hline
\textbf{\#DR} & \multicolumn{3}{c|}{735} & \multicolumn{3}{c|}{995} & \multicolumn{3}{c|}{751} & \multicolumn{3}{c|}{2481} \\
\hline
\textbf{MD} & \textbf{\#} & \textbf{\#C} & \textbf{\#V} & \textbf{\#} & \textbf{\#C} & \textbf{\#V} & \textbf{\#} & \textbf{\#C} & \textbf{\#V} & \textbf{\#} & \textbf{\#C} & \textbf{\#V} \\
\hline
\textbf{\zero{}} & 726   & 573   & 289 \tiny{(39.8\%)} & 995   & 878   & 501 \tiny{(50.4\%)} & 745   & 666   & 394 \tiny{(52.9\%)} & 2466  & 2117  & 1184 \tiny{(48.0\%)} \\
\hline
\textbf{\few{}} & 734   & 576   & 254 \tiny{(34.6\%)} & 994   & 878   & 498 \tiny{(50.1\%)} & 743   & 674   & 423 \tiny{(56.9\%)} & 2471  & 2128  & 1175 \tiny{(47.6\%)} \\
\hline
\textbf{\critique{}} & 713   & 553   & 274 \tiny{(38.4\%)} & 995   & 907   & 605 \tiny{(60.8\%)} & 751   & 670   & 429 \tiny{(57.1\%)} & 2459  & 2130  & 1308 \tiny{(53.2\%)} \\
\hline
\textbf{\expert{}} & 737   & 562   & 283 \tiny{(38.4\%)} & 995   & 864   & 500 \tiny{(50.3\%)} & 751   & 683   & 389 \tiny{(51.8\%)} & 2483  & 2109  & 1172 \tiny{(47.2\%)} \\
\hline
\textbf{\chain{}} & 726   & 581   & 393 \tiny{(54.1\%)} & 995   & 861   & 559 \tiny{(56.2\%)} & 751   & 681   & 437 \tiny{(58.2\%)} & 2472  & 2123  & 1389 \tiny{(56.2\%)} \\
\hline
\textbf{Total} & 3636  & 2845  & 1493 \tiny{(41.1\%)} & 4974  & 4388  & 2663 \tiny{(53.5\%)} & 3741  & 3374  & 2072 \tiny{(55.4\%)} & 12351 & 10607 & 6228 \tiny{(50.4\%)} \\
\hline
\end{tabular}%
}
\end{table}%


\findingbox{4}{The performance of \techname{} is sensitive to the choice of LLMs in both \techphaseA{} and \techphaseB{}.
DeepSeek-V3.2 performs best, achieving the highest MR count (\tocheck{1,246}) and applicability (\tocheck{82.9\%}) in \techphaseA{}, and showing competitive performance with GPT-4.1 in \techphaseB{}.
In contrast, GLM-4-Plus performs worst, with the lowest applicability (\tocheck{71.1\%}) and valid meta drivers (\tocheck{41.1\%}).
Considering the release dates, newer and more capable LLMs may lead to better performance.}

\subsubsection{Overview of Results}
\autoref{tab:results-by-llms-promts} presents the \techphaseA{} and \techphaseB{} results under different configurations, with different LLMs shown in the columns and different prompt strategies in the rows.
The upper subtable displays the statistics of generated MRs, whereas the lower subtable displays \techphaseB{} results. 
The Total columns report the sums of prompt strategies, while the Tot. row reports the sums by LLMs.
The \#DR row reports the number of distinct MRs gathered by each LLM; we calculate this number by aggregating all MRs generated by the same LLM and then deduplicating them.

\subsubsection{\techphaseA{} Results across LLMs}

We omit the version and use GLM, DeepSeek, and GPT to denote the studied LLMs (see \autoref{sec:llm}) for clarity. 
As shown in \autoref{tab:results-by-llms-promts}, GLM, DeepSeek, and GPT generate \tocheck{1,149}, \tocheck{1,246}, and \tocheck{1,080} MRs, of which \tocheck{71.1\%}, \tocheck{82.9\%}, \tocheck{77.3\%} are applicable.
The performance of \techname{} in \techphaseA{} varies significantly across LLMs in terms of both the number of generated MRs and the AP. 
Specifically, as the most recent model considered in our study, DeepSeek performs the best among the studied LLMs, generating \tocheck{216 (+11.8\%)} and \tocheck{198 (+5.6\%)} more applicable MRs than GLM and GPT.
GLM and GPT produce similar numbers of distinct MRs (i.e., \tocheck{735 and 751}), whereas DeepSeek again performs the best, generating about \tocheck{one third} more distinct MRs.
These results indicate that the effectiveness of \techphaseA{} is sensitive to the choice of LLM. 

\subsubsection{\techphaseB{} Results across LLMs}
We instruct each LLM to implement meta drivers based on the distinct MRs it produced. 
The purpose of this experimental setup is to evaluate each LLM's ability to handle the \mfoe{} task independently.
In practice, the MRs generated by an LLM are general and can be used by another LLM as well.
As shown in \autoref{tab:results-by-llms-promts}, GLM, DeepSeek, GPT implement \tocheck{3,636, 4,974, and 3,741} meta drivers, of which \tocheck{1,493, 2,663, and 2,072} are valid.
GLM performs the worst among the three models, with a VP of only \tocheck{41.1\%}; it cannot even stably generate meta driver candidates in a quantity that exactly matches the number of distinct MRs (\tocheck{735}), instead producing \tocheck{713\rangeline{}737} meta driver candidates.
GPT achieves the best VP, with \tocheck{55.4\%} of meta driver candidates being valid. 
Although DeepSeek implements the most valid meta drivers (\tocheck{2,663}), this can be attributed to its largest number of input MRs (\tocheck{995}), and its VP is \tocheck{1.9\%} lower than that of GPT.
These results indicate that the effectiveness of \techphaseB{} is also highly dependent on the choice of LLM.

The columns \tocheck{4\rangeline{}6} of \autoref{tab:fuzz-caps-means} report the fuzzing results achieved by the meta drivers implemented using different LLMs, where the better results are colored in darker blue.
As shown, the meta drivers implemented by GPT achieve the best performance in terms of code coverage, obtaining the highest average edge coverage on \tocheck{nine} out of \nmfoesubjects{} projects.
Regarding triggered crashes, DeepSeek performs the best, as its meta drivers trigger the highest average number of crashes on \tocheck{six} projects.
Overall, these results indicate that DeepSeek and GPT achieve comparable performance, each showing advantages in generating meta drivers that excel in  crash detection and code coverage.
Moreover, these findings confirm that \techname{} is sensitive to the choice of LLMs in the \techphaseB{} phase.

\subsection{\rqline{5}{Prompt Sensitivity}}
\label{subsec:rq-prompts}

\findingbox{5}{Compared with the sensitivity to LLM choices, \techname{} is less sensitive to the choice of prompt strategies.
All strategies perform closely in \techphaseA{}, with a standard deviation of \tocheck{0.7\%}.
For \techphaseB{}, \chain{} performs the best by implementing the largest number of valid meta drivers (\tocheck{1,389}) with the highest valid proportion (\tocheck{56.2\%}), whereas \expert{} performs the worst.
Regarding average fuzzing capabilities, \expert{} generates meta drivers that achieve the best results, where \chain{} performs the worst in both edge coverage and crash detection.
}

\subsubsection{\techphaseA{} Results across Prompts.}
The rows of \autoref{tab:results-by-llms-promts} present the results of \techname{} under different prompt strategies, where the totals are reported in the last three columns.
As shown, \techname{} achieves similar performance across different prompt strategies in terms of AP, with a standard deviation of only \tocheck{0.7\%}; and \chain{} achieves the highest AP (78.1\%).
Despite the similar AP values, \chain{} generates the largest number of MRs (\tocheck{856}), while the other four strategies generate comparable numbers of applicable MRs (\tocheck{453\rangeline{}464}).
This is because \chain{} is configured to generate MRs based on the provided project information and is not constrained by a predefined number, while the other four strategies are instructed to generate \ngenMRs{} in total (\ngenMRsEach{} each time).
We adopt this configuration because these four strategies tend to produce duplicates after generating around \ngenMRs{} MRs, while \chain{} does not exhibit this issue.
Overall, \chain{} performs the best among the five studied prompt strategies in \techphaseA{}, and \techname{} is largely insensitive to the choice of prompt strategies.

\subsubsection{\techphaseB{} Results across Prompts.}
As shown in the lower subtable of \autoref{tab:results-by-llms-promts}, \techname{} achieves VP values ranging from \tocheck{47.2\%\rangeline{}56.2\%} across the five prompt strategies, with \chain{} performing the best and \expert{} the worst. 
The columns \tocheck{7\rangeline{}11} of \autoref{tab:fuzz-caps-means} present the fuzzing results achieved under different prompt strategies, where the larger number of edges (crashes) is indicated in darker green.
As shown, the meta drivers implemented under different prompt strategies exhibit similar coverage, with \expert{} performing the best with slight overall advantages.
For crash detection, \expert{} also performs the best, generating meta drivers that trigger the highest average number of crashes on \tocheck{six} projects.
Surprisingly, \chain{} performs the worst in both edge coverage and triggered crashes, despite its superior generation statistics in terms of MRs and meta drivers.
This implies that \chain{} is effective in accomplishing the tasks of \mfoe{}, but cannot guarantee the fuzzing capabilities of meta drivers.
A mixed prompting strategy (e.g., \chain{}+\expert{}) may yield better overall results.

\section{Discussion}




This section presents (1) a case study on crash analysis and LLM-assisted false-positive mitigation, (2) lessons learned, and (3) threats to validity.

\subsection{Case Study}
\label{subsec:case-study}
\begin{figure}
    \centering
    \includegraphics[width=\linewidth]{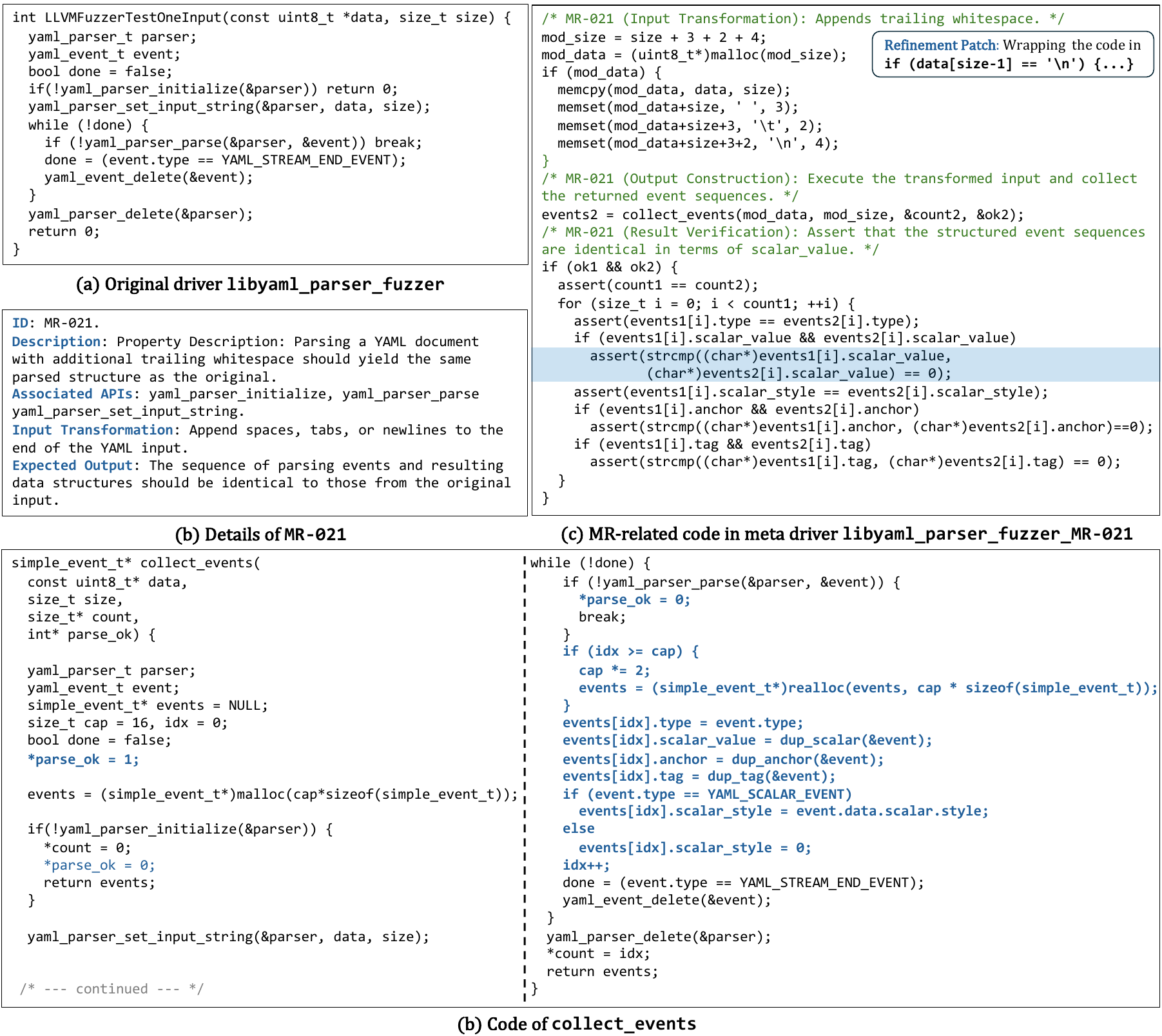}
    \caption{A meta driver implemented in Libyaml (S06).}
    \label{fig:case-study}
    \Description{Case study}
\end{figure}

\autoref{fig:case-study} illustrates a meta driver constructed in Libyaml (S06). 
The original driver is shown in the subfig (a), the implemented MR (MR-021) in (b), and the core MR-related code in the meta driver in (c).
As shown, the original driver implements a basic test sequence, which initializes the test input, invokes the core API under test (\texttt{yaml\_parse\_initialize}), and releases resources after fuzzing.
\texttt{MR-021} captures a whitespace-invariance property: appending trailing whitespace to a YAML document should not change its parsing results, i.e., both the event sequence and resulting data structures remain identical.
The meta driver incorporates \texttt{MR-021} by appending trailing whitespace (i.e., three spaces, two tabs, and four newlines) to the input, re-parsing it, and verifying the equivalence of structured event sequences between the original and transformed executions.
The core function of the meta driver is \texttt{collect\_events}, which parses a given YAML input and collects the resulting sequence of parsing events into a structured array for subsequent comparison; the source code is shown in \autoref{fig:case-study}(d), with the MR-supporting code in blue.
At first glance, \techname{} correctly implements the meta driver, with a valid MR and functional metamorphic testing logic.

\subsubsection{Meta-driver-induced False Positives}
Through \fuzzhours{}-hour fuzzing, the meta driver shown in \autoref{fig:case-study} triggers 141 crashes in total, all of which are assertion failures. 
After deduplication based on crash summaries, we observe that all crashes in fact originate from a single unique assertion failure (AF), which corresponds to a violation of the \texttt{scalar\_value} check enforced by the \texttt{assert} statement highlighted in red.
We further analyze this AF by examining the YAML specification and consulting ChatGPT, and conclude that it is a false positive.
The failure stems from an invalid MR assumption rather than a Libyaml bug. 
Appending trailing whitespace can alter the parsed content of block scalars (e.g., ``\texttt{|}'' and ``\texttt{>}'') without a terminating newline\footnote{YAML specification (block): \url{https://yaml.org/spec/1.2.1/\#Block}}
, under which \texttt{MR-021} does not hold, leading to false positives.
This observation indicates that, although LLMs are trained on large corpora and are capable of code understanding and test generation, they still cannot cover all corner cases and may therefore fail to implement fully correct metamorphic-based oracles.


\subsubsection{Mitigating False Positives}
\label{subsec:reduce-fp}
We explore mitigating false positives with the help of ChatGPT. 
To this end, we ask ChatGPT to refine the meta driver and provide it with three inputs: \lineno{1} the meta driver shown along with \texttt{MR-021}, \lineno{2} the root cause description of the false positive (i.e., how the MR may not hold for YAML block scalars), and \lineno{3} a example crashing input generated during fuzzing.
ChatGPT returns a patching plan that restricts the transformation to inputs ending with ``\texttt{\textbackslash n}'', and generates a refined meta driver. 
This ensures proper YAML termination, preventing appended whitespace from affecting scalar content and allowing \texttt{MR-021} to hold for the \texttt{assert} highlighted in blue.
The refined patch is shown in the upper-right corner of \autoref{fig:case-study}(c). 
A three-hour fuzzing of the refined meta driver still triggers \tocheck{162} crashes, all of which originate from the blue-highlighted \texttt{assert} statement. 
This result is unexpected and motivates a deeper inspection of the generated inputs.
We observe that, although false positives persist, the crashing inputs no longer exhibit the same pattern (i.e., ending with ``\texttt{>}'' or ``\texttt{|}'') as those from the original meta driver, indicating that the patch is effective.
This suggests that the false positives are only partially mitigated, with the addressed cases largely corresponding to the provided example inputs (i.e., those ending with ``\texttt{>}'' or ``\texttt{|}'').
Overall, while LLM-based refinement can mitigate false positives, it may require multiple iterations with diverse crashing inputs. We leave this for future work.

\subsection{Lessons Learned}
We summarize the following lessons that may be helpful to the fuzzing community.

\textbf{\lineno{1} Fuzz Drivers Need Functional Oracles.}
Our analysis of \nossfuzzdrivers{} shows that 81.1\% of \ossfuzz{} drivers lack functional oracles. Unlike implicit crash oracles, functional oracles require a deep understanding of the target library and are difficult to implement, especially for non-vendor users. 
We thus call for better oracle design from vendors and highlight the need for automated techniques (such as our approach \techname{}) to support fuzz oracle construction.

\textbf{\lineno{2} LLMs are Capable in \mfoe{}.}
Our experiments with \nllm{} LLMs and \nprompt{} prompt strategies demonstrate that LLMs are feasible for realizing \mfoe{}. 
Better performance can be achieved with more capable models (e.g., DeepSeek-V3.2) and mixed prompt strategies (e.g., \chain{}+\expert{}), as different prompts excel at different aspects.
We hope these findings can inform future research.

\textbf{\lineno{3} LLM-generated Meta Drivers Need Refinement.}
Although LLMs are capable of generating applicable MRs and valid meta drivers, the resulting meta drivers may still suffer from false positives even after multi-stage validation. 
Our case study shows that LLMs can provide refinement patches to mitigate such issues, suggesting the need for an additional refinement stage to iteratively improve the quality of generated meta drivers. 
We leave this to future work.

\subsection{Threats to Validity}
\subsubsection*{Internal Validity}
Internal validity correlates with the completeness and fairness of the methodology \cite{ampatzoglou2019identifying, gorz2025empirical}.
In our study, threats to internal validity mainly arise from the generation and evaluation of MRs and modified drivers.
We address these threats by \lineno{1} constructing \mfoe{} workflows based on established studies in metamorphic testing \cite{chen2018metamorphic,zhang2025can}, fuzz driver generation \cite{zhang2024effective},  and LLM for software engineering \cite{sun2025source}, and by \lineno{2} adopting and refining evaluation metrics from these pioneering works. 
These efforts help to ensure internal validity to a large extent.

\subsubsection*{External Validity}
External validity refers to the extent to which the results can be generalized to subjects beyond those included in the study \cite{ampatzoglou2019identifying, gorz2025empirical}.
In this paper, we \lineno{1} investigate various fuzz drivers from \ossfuzz{} and \lineno{2} include multiple LLMs (\autoref{sec:llm}) and prompting strategies (\autoref{subsec:prompt}).
We also perform fuzzing for \ncpuhours{} CPU hours to evaluate the capabilities of the valid meta drivers.
These efforts strengthen the external validity.

\section{Related Work}

\subsubsection*{Fuzz Driver Generation}
Fuzz drivers are essential components of fuzzing and a growing body of work has been devoted to its automatic generation \cite{jeong2023utopia,lyu2024prompt,sherman2025no,chen2023hopper}.
Recent studies continue to tackle the open challenges of automatic driver generation; examples include WildSync \cite{wu2025wildsync} and LibErator \cite{toffalini2025liberating}. 
With the advancement of modern LLMs, intelligent driver generation techniques have also emerged \cite{zhang2024effective, liu2025promefuzz}.
In contrast to prior work that focuses on synthesizing call sequences and expanding API coverage, our study shifts attention to oracles, the critical components responsible for bug detection.
Note that although \citet{sherman2025no} propose an oracle-guided technique, the ``oracle'' in there refers to evaluation criteria for driver quality rather than the bug-detection mechanism, making our study fundamentally different.


\subsubsection*{Metamorphic Testing}
Metamorphic testing is one of the most prominent techniques for alleviating the oracle problem \cite{barr2014oracle, howden2006theoretical} and has addressed intricate challenges in testing complex or non-traditional software systems \cite{le2014compiler, guo2024cootest, mu2025improving}. 
It has also attracted significant interest from the fuzzing community and has been integrated into fuzzing techniques as an extended mechanism to detect bugs beyond easily observable crashes \cite{schloegel2024sok, liu2023generation, tsigkanos2023large,pan2024edefuzz,kokkonis2025rosa}. 
In contrast to these studies, our study focuses on incorporating metamorphic-based oracles into fuzz drivers and is \textit{not} limited to specific types of bugs.
Researchers have explored the use of LLMs to support metamorphic testing.
Pioneering groups include \citet{tsigkanos2023large}, \citet{shin2024towards}, and \citet{zhang2025can}.
Inspired by this line of work, our study employs LLMs to generate MRs and construct stronger fuzz oracles, extending the pioneering efforts.




    


\subsubsection*{LLM-based Testing}
Modern LLMs have significantly reshaped software engineering (SE) \cite{he2025llm,abrahamsson2017agile,wei2024requirements}.
Many SE tasks, such as code summarization \cite{virk2025calibration,crupi2025effectiveness} and program repair \cite{xie2025premm,bouzenia2025repairagent}, have explored leveraging LLMs to improve effectiveness, intelligence, and automation. 
In addition to fuzzing and metamorphic testing, LLMs have also been applied to other fields of automatic testing \cite{harman2025harden}.
For example, \citet{shang2025large} conduct a systematic study on the adoption of LLMs in unit testing.
\citet{lingllm} employ LLMs to test report clustering and facilitate crowdsourced testing.
\citet{lemieux2023codamosa} leverage LLMs to facilitate search-based testing by escaping coverage plateaus. 
Unlike these studies, our study centers on fuzzing and complements their efforts. 

\section{Conclusion}
In this paper, we investigate current practices in fuzz oracle implementation by analyzing \nossfuzzdrivers{} drivers collected from \ossfuzz{}.
Our results show that \tocheck{81.1\%} of the drivers lack functional oracles for bug detection, highlighting the need for stronger fuzz oracles.
Motivated by this observation, we develop an LLM-based pipeline that leverages modern LLMs to augment existing fuzz drivers with metamorphic-based oracles.
Through systematic experiments with \nllm{} LLMs and \nprompt{} prompting strategies, we generate \tocheck{2,685} applicable MRs and \nvmd{} valid meta drivers, demonstrating the feasibility of metamorphic-based fuzz oracle enhancement.
A fuzzing exceeding \ncpuyears{} CPU years of the meta drivers shows that the meta drivers improve edge coverage by \tocheck{18.7\%} over the original drivers and detect \tocheck{1,528} unique crashes.
Finally, we perform a case study to analyze the detected crashes, and distill actionable insights for the fuzzing community.


\section{Data-availability Statement}
We publish our dataset, scripts, and experimental results in this repository \cite{techArtifact}.

\bibliographystyle{ACM-Reference-Format}
\bibliography{bibfile}

\end{document}